\documentclass[prd,
twocolumn,
eqsecnum,floatfix,letterpaper,superscriptaddress,nofootinbib]{revtex4}
\usepackage[utf8]{inputenc}
\usepackage{color} \usepackage[normalem]{ulem}
\usepackage{amsmath,amssymb,graphicx} \usepackage{bm} \usepackage{times}
\usepackage{microtype} \usepackage{csquotes} \usepackage{booktabs}
\newcommand{\e}{\mathrm{e}} \usepackage{subfigure} \usepackage[normalem]{ulem}
\usepackage{setspace}

\usepackage[varg]{txfonts} \usepackage[colorlinks, pdfborder={0 0 0}]{hyperref}
\usepackage{natbib}
\definecolor{LinkColor}{rgb}{0.75, 0, 0} \definecolor{CiteColor}{rgb}{0, 0.5,
0.5} \definecolor{UrlColor}{rgb}{0, 0, 0.75} \hypersetup{linkcolor=LinkColor}
\hypersetup{citecolor=CiteColor} \hypersetup{urlcolor=UrlColor}
\begin{document}
\title{Spin-induced deformations and tests of binary black hole nature using third-generation detectors}
\date{\today} 
\author{N. V. Krishnendu} \email{krishnendu@cmi.ac.in}
\affiliation{Chennai Mathematical Institute, Siruseri, 603103, India.}
\author{Chandra Kant Mishra}\email{ckm@iitm.ac.in} 
\affiliation{Indian Institute of Technology Madras, Chennai, 600036, India.}
\author{K. G. Arun} \email{kgarun@cmi.ac.in}
\affiliation{Chennai Mathematical Institute,  Siruseri, 603103, India.} 
\begin{abstract}
In a recent {\it letter}  [N. V. Krishnendu {\it et al}., Phys. Rev. Lett. 119, 091101 (2017)]  we explored the possibility of probing the binary black hole nature of
coalescing compact binaries, by measuring their spin-induced multipole moments, 
observed in advanced LIGO detectors. Coefficients characterizing the
spin-induced multipole moments of Kerr black holes are predicted by the
``no-hair" conjecture and appear in the gravitational waveforms through
quadratic and higher order spin interactions and hence can be directly measured
from gravitational wave observations. By employing a non-precessing
post-Newtonian (PN) waveform model, we assess the capabilities of the
third-generation gravitational wave interferometers such as Cosmic Explorer and
Einstein Telescope in carrying out such measurements and use them to test the
binary black hole nature of observed binaries. In this paper, we extend the
investigations of [N. V. Krishnendu {\it et al}., Phys. Rev. Lett. 119, 091101 (2017)],
limited to measuring the binary's spin-induced quadrupole moment using their
observation in {\it second} generation detectors, by proposing to measure (a)
spin-induced {\it quadrupole} effects using {\it third} generation detectors,
(b) {\it simultaneous} measurements of spin-induced {\it quadrupole} and {\it
octupole} effects, again in the context of the third-generation detectors. We
study the accuracy of these measurements as a function of total mass, mass
ratio, spin magnitudes, and spin alignments. Further, we consider
two different binary black hole populations, as proxies of the population that will be observed by the third generation detectors, and obtain the resulting distribution of the spin-induced quadrupole coefficient. This helps us assess how common are those cases where this test would provide very stringent constraints on the black hole nature. These error bars provide us upper
limits on the values of the coefficients that characterize the spin-induced
multipoles. We find that, using third-generation detectors the symmetric
combination of coefficients associated with the spin-induced quadrupole moment of
each binary component may be constrained to a value $\leq 1.1$ while a similar
combination of coefficients for spin-induced octupole moment may be constrained to $\leq 2$, where both combinations take the value of 1 for a binary black hole system. These estimates suggest that third-generation detectors can
accurately constrain the first four multipole moments of the compact objects
(mass, spin, quadrupole, and octupole) facilitating a thorough probe of their
black hole nature.
\end{abstract}

\maketitle
\section{Introduction}
\label{sec:intro}
Recent detections of binary black hole (BBH) mergers by Laser Interferometric
Gravitational wave Observatory (LIGO) and VIRGO gravitational wave observatory have confirmed the existence of binary black holes in nature and that they merge under the effect of gravitational wave (GW) radiation reaction \cite{Discovery, GW151226,GW170104,GW170608,GW170814,LIGOScientific:2018mvr,LIGOScientific:2018jsj}. Several tests of general relativity \cite{Meidam:2014jpa,Yunes:2013dva,AIQS06a,AIQS06b,Berti:2018cxi, Berti:2018vdi,Will:1977wq,TIGER2014,Yunes:2009ke,Will:1997bb,Samajdar:2017mka, Ghosh:2017gfp} were performed using these signals leading to the first ever bounds on potential deviation from  the theory in the strong-field regime of gravity \cite{TOG,O1BBH,GW170104,GW170608,GW170814}. These tests make use of the fact that the dynamics of the compact binary, and hence the gravitational
waveform could be different in an alternative theory of gravity. Hence the
observation of binary black holes can lead to constraints on possible
departures from general relativity.

The binary black hole dynamics consists of three major phases inspiral, merger
and ringdown. One can model the inspiral phase using post-Newtonian formalism\cite{Blanchet} whereas numerical relativity simulations are needed to model the merger regime \cite{Pretorius:2007nq}. In order to study the ringdown part of the dynamics, one may use black hole perturbation theory techniques
\cite{Sasaki:2003xr}.  While the observations to date are consistent with
this binary black hole dynamics, there may still be room for explaining these
observed mergers as due to mergers of some exotic compact
objects~\cite{Giudice}.  These exotic compact objects could mimic the
properties of the black holes up to the accuracy with which we are currently
able to extract the signal and its parameters. As the gravitational wave interferometers become more and more sensitive, our parameter estimation
accuracies should improve dramatically enabling a thorough probe of the nature
of these compact objects. Proposed third-generation ground-based detectors such as Einstein Telescope (ET-D) \cite{Sathyaprakash:2011bh} and Cosmic Explorer (CE) \cite{Regimbau:2012ir, Hild:2010id, Hild:2008ng} hence have the strong potential to probe the nature of compact binaries which motivates this work.
	
Leading candidates for these black hole mimickers include gravastars
\cite{gravastar_Mazur}, boson stars \cite{Bosonstar_Liebling} and firewalls
\cite{Firewall}. Modeling mergers of these exotic objects is a hard problem and
their direct deployment for data analysis is not likely to happen in the near
future. So a more pragmatic approach would be to devise tests which are generic and model independent and are based on our solid understanding of the binary black hole dynamics. An efficient method to probe the presence of exotic
compact objects is to understand the possible ways in which such objects can
correct for the properties of black holes which can be detected or ruled out by
introducing appropriate free parameters in the gravitational waveform. These
tests are often referred to as ``null tests" as the free parameters are zero
for binary black holes. In order to develop such model independent null tests
of black hole mimickers, it is important to identify those properties which are
unique to black holes and trace their imprints on the gravitational waveform so
that we can measure them from observations.

One of the characteristic properties of black holes in the general theory of
relativity is related to the ``no-hair" conjecture, which says that all the
multipole moments of a Kerr black hole are completely specified by its mass and
spin. This means that, it is always possible to relate the $\ell^{\rm th}$
multipole of the Kerr black hole to the mass ($M$) and the dimensionless spin
parameter ($\chi=S/M^2$) as, ${M}_{\ell}$ + $\rm{i}\,{S}_{\ell}$ =
${M}^{\ell+1} (\rm{i}\,{\chi})^{\ell}$ \cite{Hansen74, Carter71,
Gurlebeck:2015xpa,Ryan:1995wh,Geroch2,Geroch1}. Here ${M}_{\ell}$ and
$S_{\ell}$ are the mass- and the current-type multipole moments,
respectively. This property leads to several observational predictions unique
to a black hole which are built-in to the gravitational waveform facilitating
tests of black hole nature, some of which are discussed below.
\subsection{Tests of binary black hole nature using gravitational waves} The
fact that a black hole cannot be tidally deformed, leads to a vanishing tidal
Love number~\cite{RelativistictheoryofTLNs_EricBinnington,NHTBHsAstrophysicalEnvironments2015}. Using a gravitational wave phasing formula which contains the tidal Love numbers~\cite{FlanaganHindererNSTLNs, Vines}, one can directly measure these parameters from observations which in turn can be used to constrain the nature of the compact object constituting the binary system \cite{Cardoso:2017cfl, Sennett:2017etc, Johnson-McDaniel2018}. Measurement of tidal deformability parameter from gravitational wave observations for various neutron star models is also studied in different contexts \cite{FlanaganHindererNSTLNs}.  Recently,  Cardoso {\it et al.}~\cite{Cardoso:2017cfl} have calculated the tidal deformability parameters of non-black hole compact objects (including boson stars, gravastars,
wormholes, and other toy models for quantum corrections at the horizon scale)
and have studied the detectability of such parameters using advanced
gravitational wave detectors.  In Ref.  \cite{Sennett:2017etc}, authors
studied the distinguishability of boson star systems from black holes and
neutron stars by measuring the tidal deformability parameter. A rigorous
formulation of this test using Bayesian inference~\cite{Johnson-McDaniel2018}
has brought the idea closer to be implemented on detected gravitational wave
events. 

Another way to test the black hole nature is by using the quasi-normal
modes~\cite{VishuNature} of the  perturbed black hole formed by the
merger~\cite{BHspect04,Berti:2009kk,Meidam:2014jpa,Berti:2009kk}. For a Kerr
black hole, all the quasi-normal modes are characterized by the mass and spin
of the black hole according to the \enquote{no-hair} conjecture.  Though the
waveform models for exotic compact objects are less developed,  there have been various attempts to calculate the quasi normal modes of boson
stars~\cite{Macedo:2016wgh,BertiBS,PhysRevD.88.064046} and
gravastars~\cite{0264-9381-24-16-013,QNMsGSsPaniAxialPolar,QNMsGSsChirenti2017}.
These can be used to discern boson stars and gravastars from black holes.

Measurement of the so called tidal heating parameter can also be used as a tool
to test the black hole nature. Consider a black hole event horizon surrounded
by external gravitating objects. The rotational energy of this black hole may
dissipate gravitationally due to the tidal disruption of exterior matter
\cite{Hartle:1973zz}. The loss of energy and angular momentum of a Kerr black
hole near the horizon can lead to non-zero values of the tidal heating
parameter. The measured value of the tidal heating parameter will be zero for
any system without an event horizon.  The tidal heating effect shows up in the
gravitational wave phasing~\cite{Katerina2012, Katerina2016} which helps us to
measure this effect from observations~\cite{Maselli:2017cmm} and thereby test
the black hole nature of the compact object.
 
It has been found that the multipole moment structure of a central compact
object can be extracted from the dynamics of a less massive object orbiting
it~\cite{PhysRevD.56.1845,Rodriguez:2011aa,Brown:2006pj}.  Reference
\cite{2004PhRvD..69l4022C} introduced the \enquote{bumpy black holes} as a
model of space-times which deviate from that of Kerr black holes. Bumpy black
holes and their astrophysical importance is extensively studied
in~\cite{2006CQGra..23.4167G}. 

Recently, Ghosh {\it et. al} proposed a method \cite{Ghosh:2017gfp,
2018CQGra..35a4002G} to study the consistency of the inspiral-merger-ringdown dynamics of a binary black hole system to the one predicted by general relativity. The idea here is to infer the mass and spin parameters of the
merger remnant from the post-merger part of the gravitational wave signal and
ask if this is consistent with the same as inferred from the inspiral part of
the gravitational wave signal (using the numerical fitting formula given in
\cite{PhysRevD.90.104004}). This method allows one to quantify how close the
observed high mass compact binary mergers are to the mergers of binary black
holes in general relativity~\cite{Discovery, GW170104}. 

\begin{figure}[hbtp]
\includegraphics[scale=0.55]{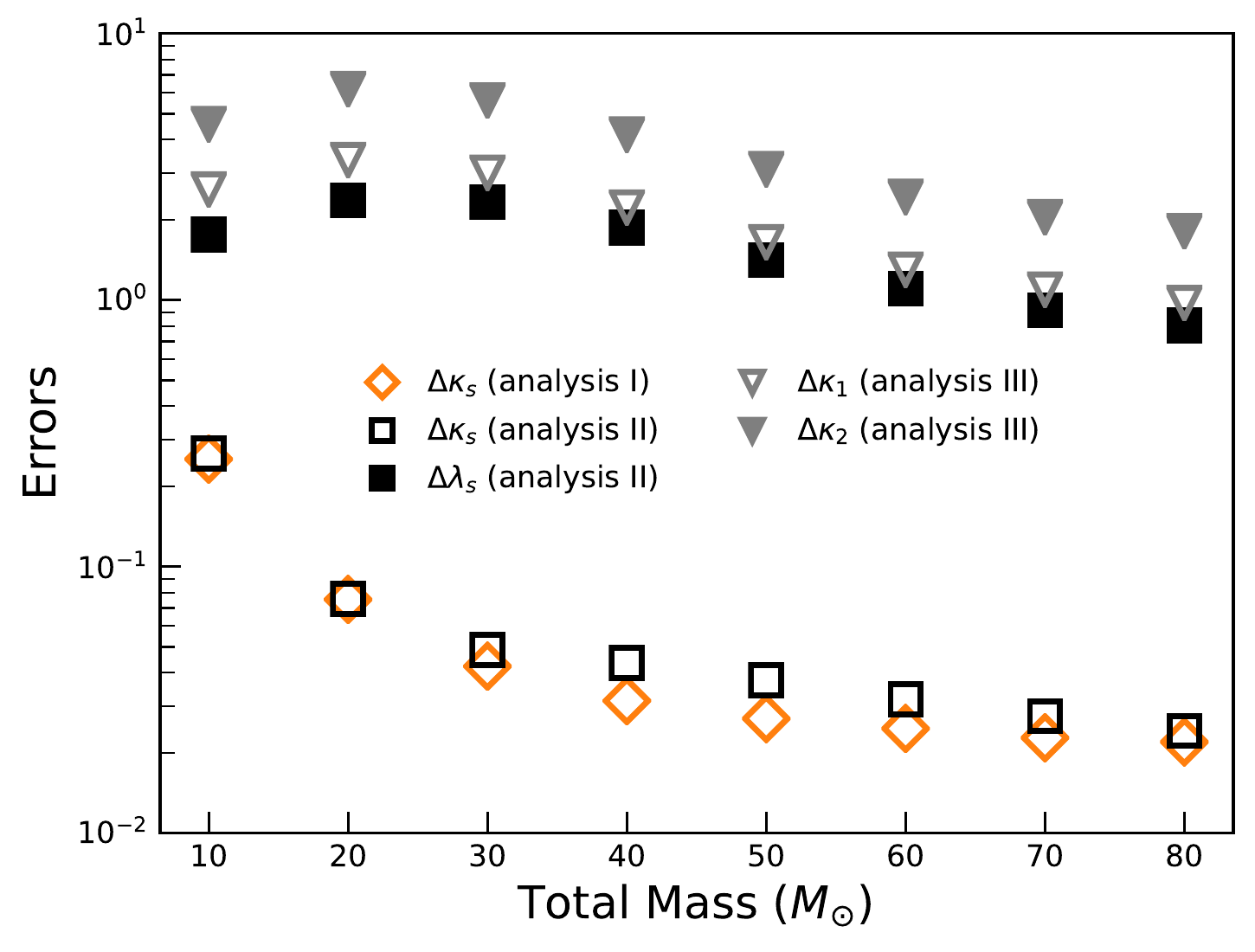}
\caption{Figure displays variation of $1$-$\sigma$ errors in the measurement of
		parameters characterizing spin-induced multipole moments as a function of the total mass of the binary for the three different analyses.  {\bf Analysis I} represents the case where $\kappa_s=(\kappa_1+\kappa_2)/2$ is treated as an independent parameter (here $\kappa_{1,2}$ are parameters characterizing the spin-induced quadrupole moment of each binary component) while the antisymmetric combination of $\kappa_1$ and $\kappa_2$ as well as the symmetric and antisymmetric combination of parameters characterizing the spin-induced octupole moment, ($\lambda_1$, $\lambda_2$), are set to their binary black hole values of ($0, 1, 0$), respectively.  In {\bf Analysis II},  both $\kappa_s$ and $\lambda_s=(\lambda_1+\lambda_2)/2$ are measured simultaneously while the antisymmetric combination $\kappa_a=(\kappa_1-\kappa_2)/2$ and $\lambda_a=(\lambda_1-\lambda_2)/2$ are set to their binary black hole values of $0$. Finally in {\bf Analysis III}, we obtain errors on $\kappa_1$ and $\kappa_2$ while keeping $\lambda_1$ and $\lambda_2$ to their BH values of $1$. The binary is assumed to be at a distance of $400$Mpc and is optimally oriented. The binary's mass-ratio is $1.2$ and posses spins of $0.9$ and $0.8$ respectively for heavier and lighter components, respectively.} 
\label{SummaryPlot}
\end{figure}

\subsection{Current work}

\begin{figure}
\includegraphics[scale=0.5]{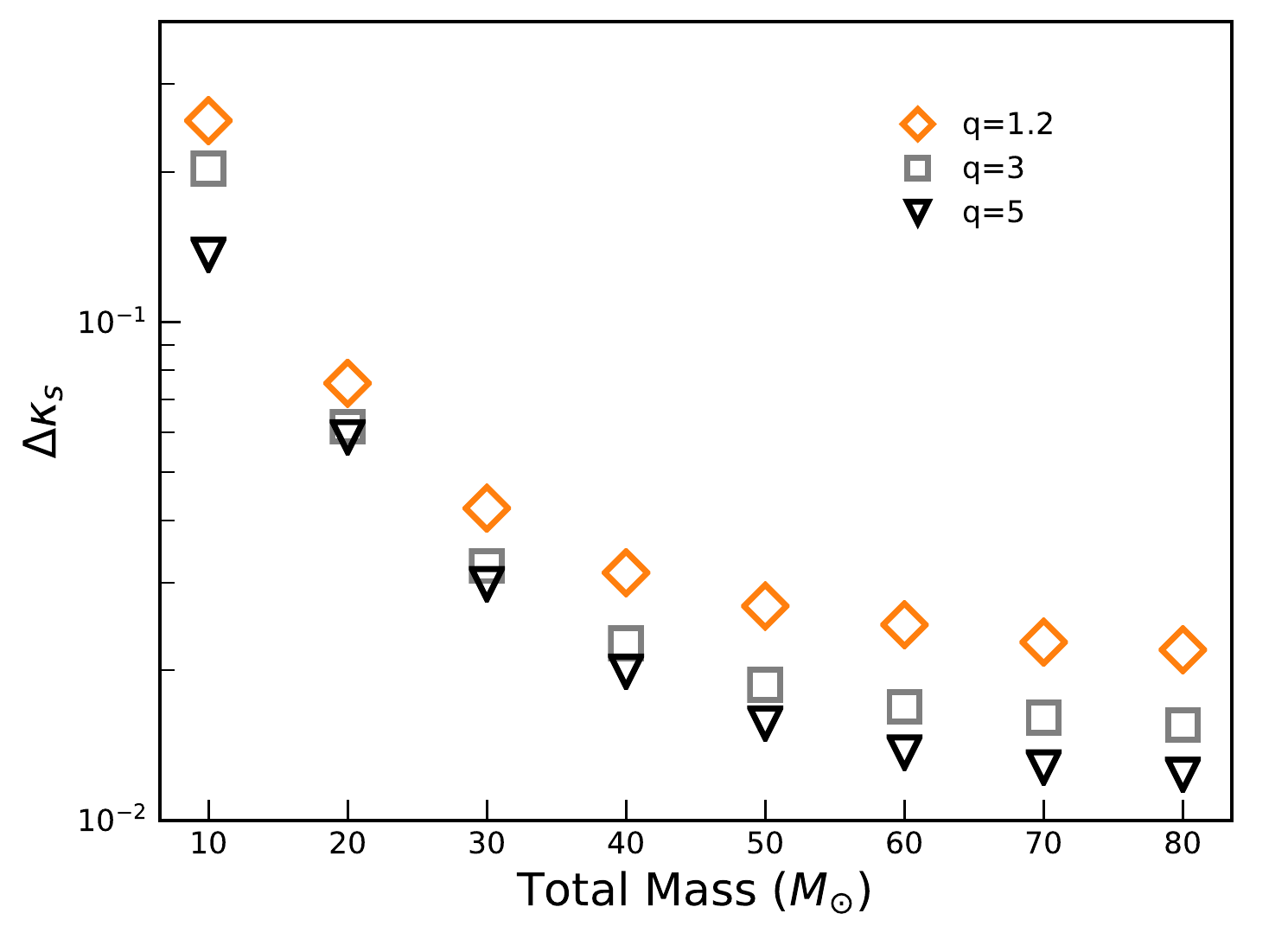}
\includegraphics[scale=0.5]{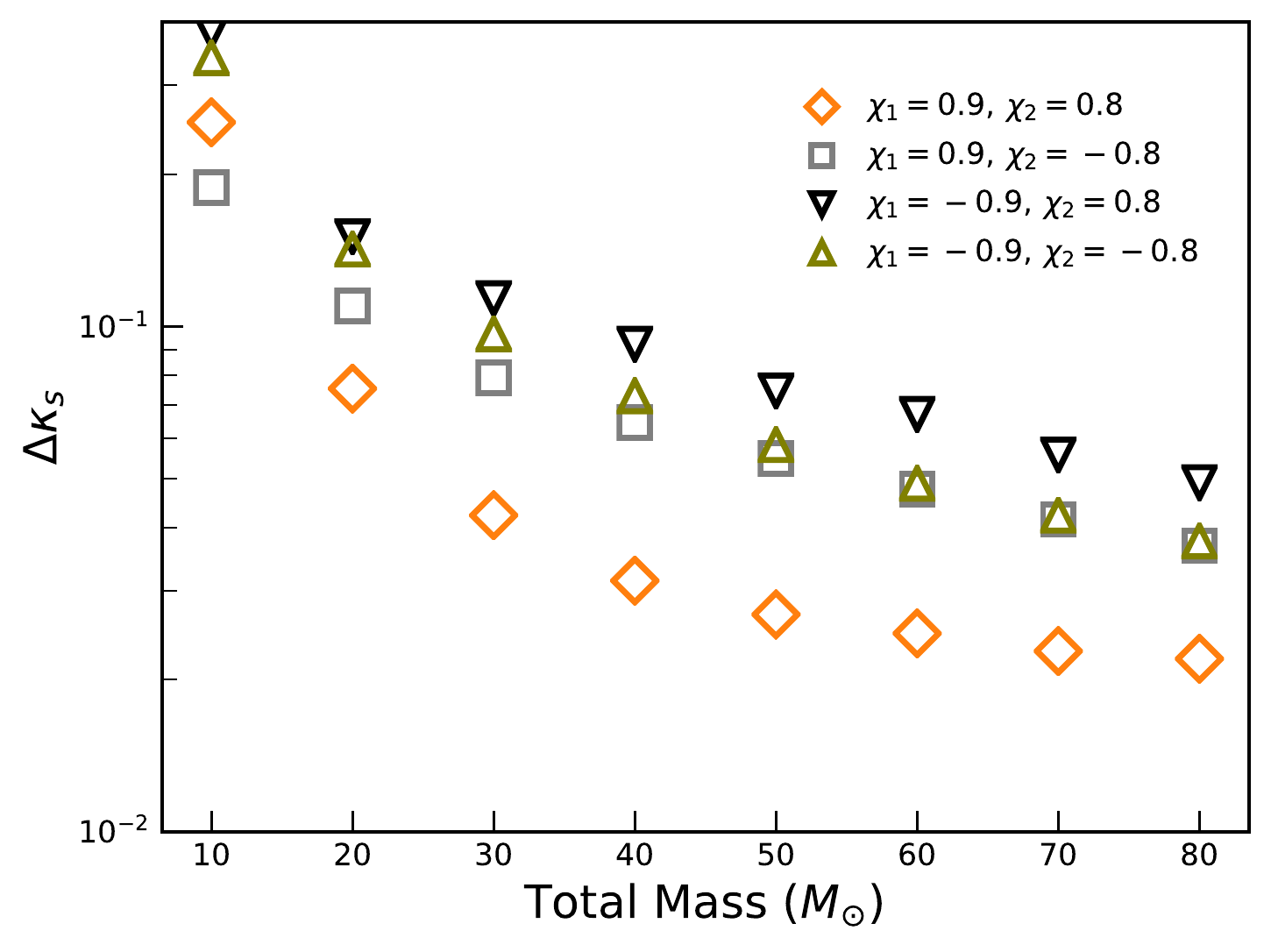}
\caption{Figure displays variation of $1-\sigma$ errors on
		$\kappa_s={(\kappa_1+\kappa_2)/2}$ (where $\kappa_{1,2}$ are parameters
		characterizing the spin-induced quadrupole moment of each binary
		component)  as a function of the binary's total mass for three
		representative mass-ratio cases with fixed component spins
		($\chi_1,\chi_2$) of $(0.9, 0.8)$ (top panel) and four representative
		spin configurations with fixed mass-ratio (q) of $1.2$ (bottom panel).}
\label{Ks_DiffMassRatios_SpinOri} \end{figure}

Recently we proposed a new method to test the binary black hole nature of
coalescing compact binary systems observable by ground-based and space-based gravitational wave interferometers \cite{Krishnendu:2017shb}. The method relies on measuring the spin-induced quadrupole moments of the binary constituents, which appear explicitly in the gravitational waveforms.  For instance, the spin-induced quadrupole moment is given by ${M}_{2}$=-${\kappa} \,{\chi}^{2} \:{M}^{3} $ where $M$ and $\chi$ are the mass and dimensionless spin parameter of the black hole  and the coefficient $\kappa$, which is a measure of the spin-induced quadrupole moment, is unity for Kerr black holes, whereas it can take values roughly between $\sim2-14$ for neutron stars
\cite{Laarakkers:1997hb,Pappas:2012qg, 2012PhRvL.108w1104P} and between
$\sim 10-150$ for boson stars \cite{Ryan97b}.  Hence an accurate and independent measurement of this coefficient for each of the binary constituents can tell us if they are indeed black holes \cite{Krishnendu:2017shb}.  For this purpose, we employed the post-Newtonian (PN) waveforms for spinning compact binaries which are explicitly parametrized in terms of these coefficients (see Sec.~\ref{sec:waveformmodel} for more details).

It was argued in Ref.~\cite{Krishnendu:2017shb} that it would not be possible
to accurately measure the deformability coefficients associated with each
binary constituents ($\kappa_1, \kappa_2$) simultaneously due to the inherent
degeneracies between them. However, the symmetric combination of the two,
$\kappa_s=(\kappa_1+ \kappa_2)/2$, can be measured accurately assuming the anti-symmetric combination is zero (which would mean that we work with the
condition $\kappa_1=\kappa_2$).  Since $\kappa_1=\kappa_2=1$ for a Kerr black hole (and hence $\kappa_s=1$ for a binary black hole), an accurate measurement of $\kappa_s$ is an excellent test of the binary black hole nature of the observed compact binary. If the binary system comprises of exotic compact objects, the measurement of the symmetric combination $\kappa_s$ should be sensitive to such a deviation from binary black hole
nature even if $\kappa_1\neq \kappa_2$. However, a further analysis, where both $\kappa_1$ and $\kappa_2$ are simultaneously measured,  will be
necessary to further understand the composition of the binary and
detailed nature of the binary constituents. This possibility is
further discussed in Sec.~\ref{sec:k1k2constraints}. The error bars associated with the measurement provides the upper limit on the
value of $\kappa_s$ allowed by the data for black hole mimicker models. These
bounds, therefore, can be mapped on the parameter space of various black hole
mimicker models. A statistically significant detection of $\kappa_s\neq 1$
could be an indication of the presence of exotic physics in play and may be
followed up. 

\begin{figure}[hbtp]
\includegraphics[scale=0.5]{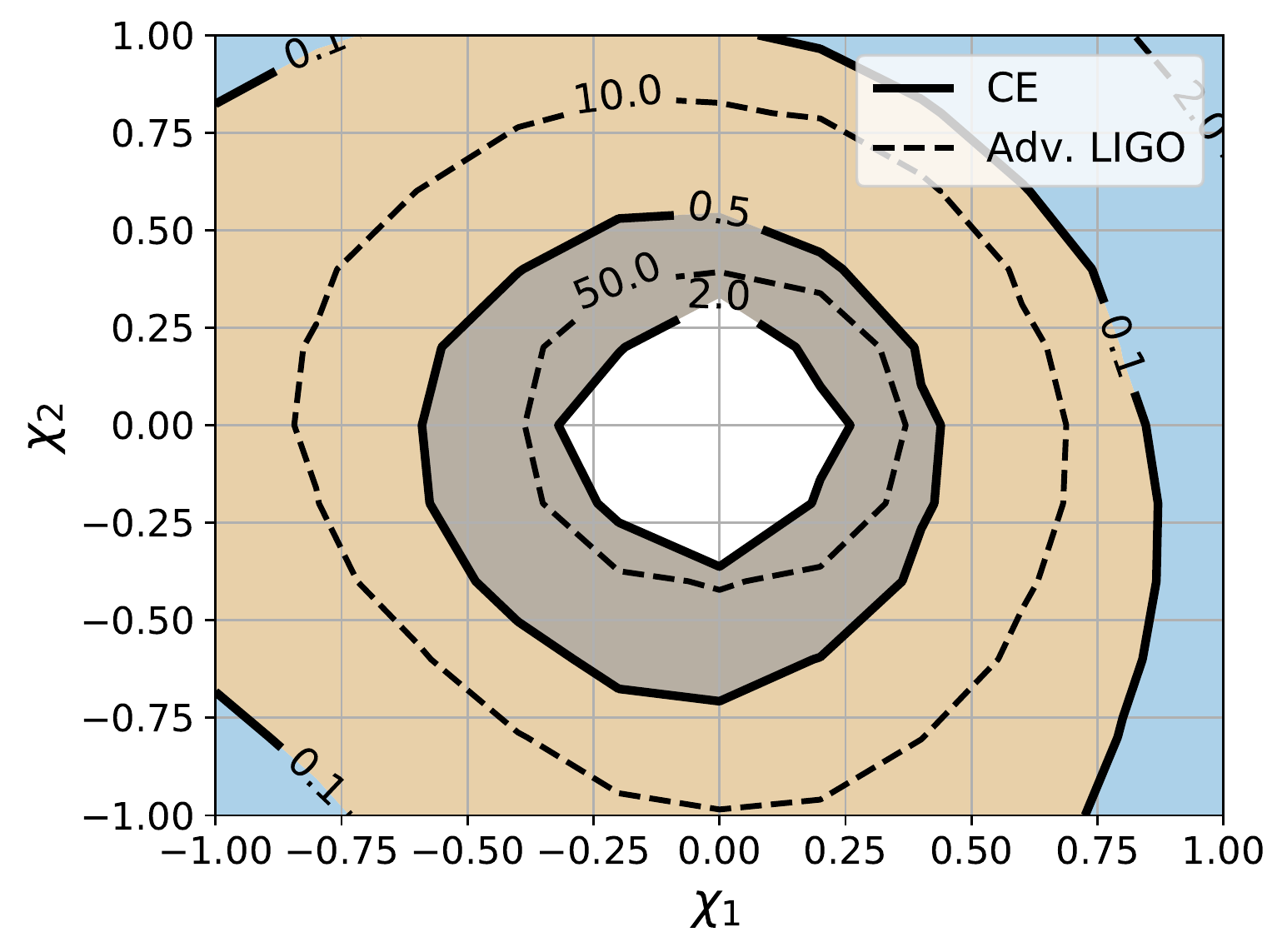}\\
\includegraphics[scale=0.5]{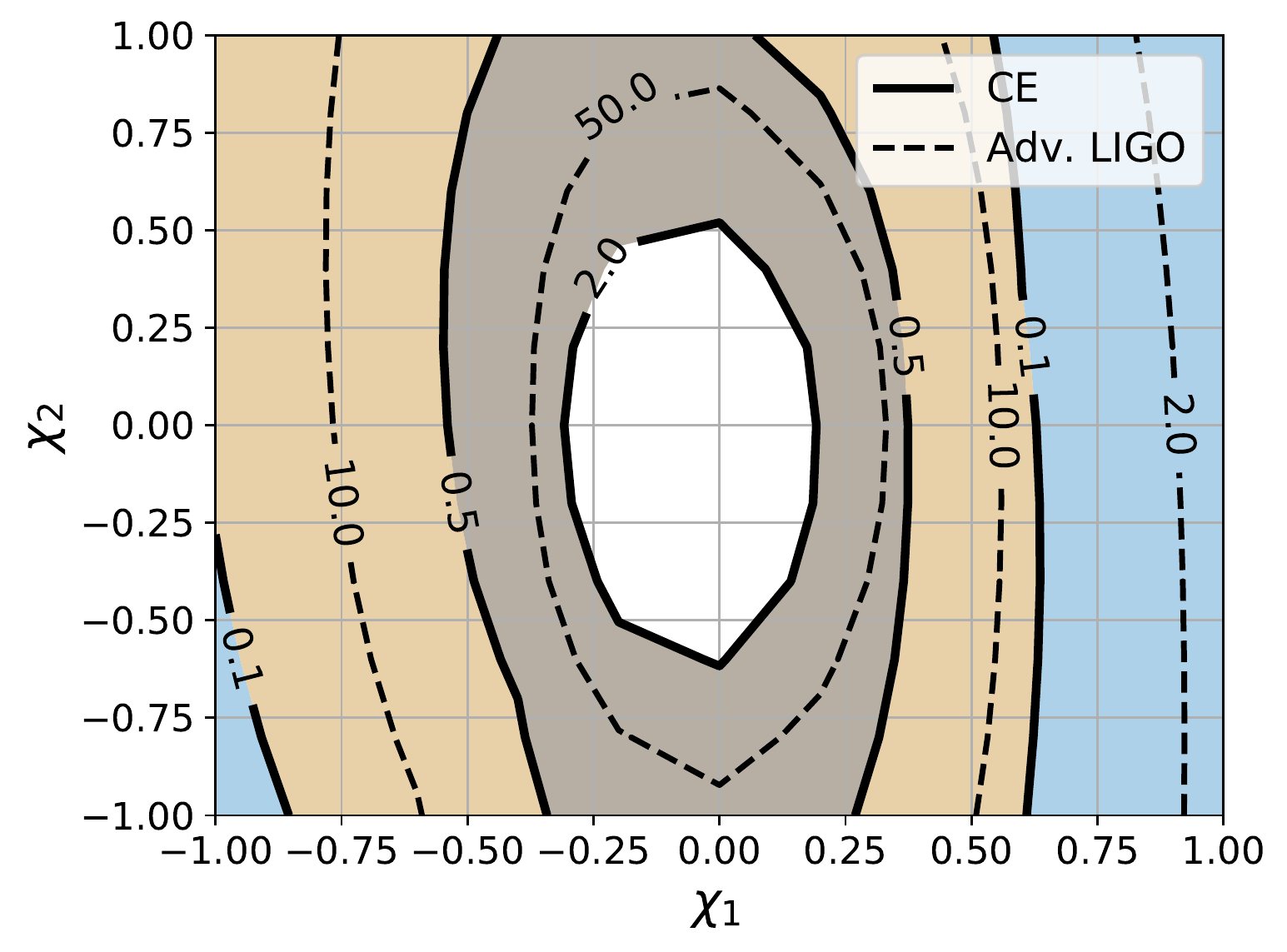} 
\caption{The errors on $\kappa_s$, the symmetric combination of $\kappa_1$ and
		$\kappa_2$, in the dimensionless spin parameter plane for the binary system with
		total mass of $30M_{\odot}$ and mass-ratios of $q=1.2$ (top panel) and
		$q=3$ (bottom panel). We assume the binary to be optimally oriented at a luminosity distance of 400Mpc. In both panels, the solid curve corresponds to the
		errors using Cosmic Explorer noise PSD and the errors using advanced LIGO noise PSD is
		denoted by dashed contours. As can be seen from the plots, parameter
		space explored in the $\chi_1$-$\chi_2$ plane is much larger for Cosmic
		Explorer compared to advanced LIGO.} 
\label{Ks_7by7_AdvLIGO_CE}
\end{figure}

In the present work, we extend the idea of \cite{Krishnendu:2017shb} in {\it
three} ways by utilizing the enhanced sensitivity of third-generation detectors
\cite{CEDwyer, Sathyaprakash:2011bh}. Firstly, we estimate the errors on
$\kappa_s$ assuming a third-generation noise sensitivity and find that the
enhanced sensitivity of third-generation detectors over second-generation
detectors improve the $\kappa_s$ estimates, roughly,  by  an order of magnitude
(see Fig.~\ref{Ks_7by7_AdvLIGO_CE}). Secondly, we investigate the ability of
third-generation detectors to simultaneously measure $\kappa_s$ and $\lambda_s$
(symmetric combination of coefficients associated with spin-induced octupole of
each binary component ($\lambda_1$, $\lambda_2$)) while we set the
anti-symmetric combinations of each pair of coefficients, ($\kappa_1$,
$\kappa_2$) and ($\lambda_1$, $\lambda_2$) to zero. This  would allow
simultaneous measurement of the mass, spin, quadrupole and octupole moments of
the source thereby permitting consistency tests between them as tests of the binary black hole
nature. Thirdly, we obtain the projected bounds on ${\kappa}_{1}$ and
${\kappa}_2$ simultaneously using third-generation detectors (keeping the
octupole moment coefficients to their BH values). These bounds can
straightforwardly be mapped to the black hole nature of the compact object
constituting the binary system leading to a much stronger test compared to the
one proposed in \cite{Krishnendu:2017shb}.
	
A summary of our analysis is shown in Fig.~\ref{SummaryPlot}, where the
projected errors on the measurement of the spin-induced multipole moments for
the three scenarios discussed above are shown as a function of total mass for a
fixed mass-ratio of $1.2$ and dimensionless spin parameters $(0.9, 0.8)$. The
binary is assumed to be optimally oriented at a luminosity distance of
$400$Mpc. The projected bounds on the binary black hole nature range from $1$
to about $8$ for the choice of mass-ratio and spin values depending on the type
of test performed. We see in Fig.~\ref{SummaryPlot} that $\kappa_s$, whether
measured alone (Analysis I) or together with $\lambda_s$ (Analysis II)  is
measured with the smallest errors. We also note that the addition of $\lambda_s$ to
the parameter space does not affect the errors on $\kappa_s$ as they are
relatively less correlated because of the different PN orders at which they
appear unlike  $\kappa_1$ and $\kappa_2$ which are strongly correlated as they
occur together in the phasing. 

The rest of the paper is organized in the following way. In
Sec.~\ref{sec:waveformmodel}, we review the idea of spin-induced multipole
moments of compact binary system within the post-Newtonian (PN) formalism. We
will briefly describe the aspects of the Fisher information matrix in
Sec.~\ref{sec2:PE}. Section~\ref{sec:results} reports the results in
detail. We conclude with Sec.~\ref{sec:conclusion}.

\section{Spin-induced multipole moments in the post-Newtonian waveforms}
\label{sec:waveformmodel}
Evolution of a compact binary system during the inspiral phase is accurately
modeled by the post-Newtonian formalism (see \cite{Blanchet:2013haa} for a
review). While sufficiently accurate post-Newtonian gravitational waveforms
(for the purposes of detection and the parameter estimation) from compact
binaries with non-spinning constituents in quasi-circular orbits were made
available as early as the early 2000s \cite{BDEI04, BFIJ02, BDIWW95}, higher order
spin effects were included through a number of recent investigations
\cite{Marsat:2012fn, Bohe:2012mr, Bohe:2013cla, Marsat:2013caa, Bohe:2015ana,
Marsat:2014xea, Arun:2008kb, Kidder:1995zr,Will:1996zj, Buonanno:2012rv,
Mishra:2016whh}. For our purposes, we choose to work with a frequency domain
waveform where the spins are (anti-) aligned with respect to the orbital
angular momentum \cite{Mishra:2016whh}. The state-of-the-art frequency domain
waveform for compact binaries with (anti-) aligned spin components incorporates
spin-orbit effects in phasing up to 4PN (leading effect appears at 1.5PN order
in the phase), spin-spin effects up to 3PN (starting at 2PN) and the leading
cubic-spin terms at 3.5PN. Moreover, the amplitude involves spin effects up to
2PN. 

The waveform we use for our analyses contain only the leading (second) harmonic (quadrupolar mode) and its PN corrections in the amplitude, while the presence of higher modes in the waveform is neglected, and schematically reads as,
\begin{equation} \tilde{h}(f)=\frac{M^2}{D_L} \sqrt{\frac{5\,\pi
\,\eta}{48}}\sum_{n=0}^{4} V_2^{n-7/2}\,C_{2}^{(n)}\,\mathrm{e}^{\mathrm{i}
\,\big({2\,\Psi}_\mathrm{SPA}(f/2)-\pi/4\big)}\, , \label{eq:waveform1}
\end{equation}
where $M$, $\eta$ and $D_L$ denote the total mass, symmetric mass-ratio and the
luminosity distance to the binary system respectively. Coefficients
$C_{2}^{(n)}$ represent the amplitude corrections to the quadrupolar harmonic
at  ({n/2}) PN order~\cite{Arun:2008kb}. The pre-factor  $V_{2}$ related to
the gravitational wave frequency $(f)$ and the total mass of the binary system
as, $V_{2}=(\pi\,M\,f)^{1/3}$. Here $\Psi_\mathrm{SPA}(f)$ represents the phase
of the waveform. Each of these $C_2^{(n)}$ and the  phasing, with explicit
dependence on spin-induced quadrupole  (through $\kappa_s$ and $\kappa_a$) and
octupole (through $\lambda_s$ and $\lambda_a$) moment parameters at respective
PN orders are given in supplemental material of~\cite{Krishnendu:2017shb}.

Effect of the leading spin-induced multipole moment (mass-type quadrupole,
$M_{2}$= $-{M}^3 \,\chi^2$) in the phasing of gravitational waves from binary
black hole systems was first computed in ~\cite{Poisson:1997ha} and contributes
to the gravitational wave phase at 2PN order. Here, the symbols $M$ and $\chi$
again represent the mass and dimensionless spin parameter for {\it each} binary
component while the negative sign (by convention) indicates that the spin
induces oblateness to the black hole.  Post-Newtonian corrections to this at
3PN order has been computed in \cite{Bohe:2015ana}. The sub-leading,
spin-induced multipole moment (current-type octupole, $S_{3}=-M^4\,\chi^3$)
starts to contribute to the phase at 3.5PN order and was computed in
\cite{Marsat:2014xea}. Notice, the spin dependences of the spin-induced
multipole moments here: $M_2 (S_3)$ have quadratic (cubic) dependences on the
spin parameter and first appear in the phasing formula at 2PN (3.5PN) order
because these are the orders at which quadratic-in-spin (cubic-in-spin) terms
start to appear in the gravitational wave phase. 

Note that the relations for $M_2$ and $S_3$ assume that the binary constituents
are black holes but can be generalized for a non-BH compact object by
introducing coefficients that characterize the degree of deformation. For
instance, we can rewrite these relations as : $M_{2}$=$-\kappa\;
M^3\,\chi^2$ and $S_{3}=-\lambda\,M^4\,\chi^3$ where the coefficients $\kappa$
and $\lambda$ take the value unity for Kerr black holes whereas they deviate from
unity for other types of compact objects including exotic alternatives to black
holes. For example, the values of $\kappa$ and $\lambda$ for neutron stars,
depending upon the neutron star equation of state and mass, range between
$\sim 2-14$ and $\sim 4-30$, respectively~\cite{Laarakkers:1997hb, Pappas:2012qg,
2012PhRvL.108w1104P}. The spin-induced multipole moments of a few exotic
compact objects  are also computed in the literature: for a particular class of
spinning boson star system $\kappa$ ($\lambda$) can take values between
$\sim\,10-150$ ($\sim \,10-200$) \cite{Ryan97b}. Variation of quadrupole and
octupole moment parameters in the boson star mass-spin parameter plane is shown respectively in Figs.~4 and 5 of \cite{Ryan97b}. Similar computations
have been done for gravastars, see for instance Refs.~\cite{Gravastars,
Uchikata:2016qku,Uchikata:2015yma} which discuss spin-induced multipole moments for thin shell gravastar models. If the observed values of spin-induced quadrupole moments
are offset from black hole value, it may be interpreted as an evidence of an
exotic compact object. On the other hand, if the posterior distribution for
the observed value is found to be peaking at 1 with a width, the corresponding
error bars can be translated into an upper bound on the allowed value of the
parameter for the particular system. In this work, we compute the projected
accuracies on the measurement of the spin-induced multipole moments using the semi-analytical parameter estimation technique of the Fisher information matrix. The necessary details of the scheme and the analysis are presented in the next
section.
\begin{figure}
\begin{center}
	\includegraphics[scale=0.55]{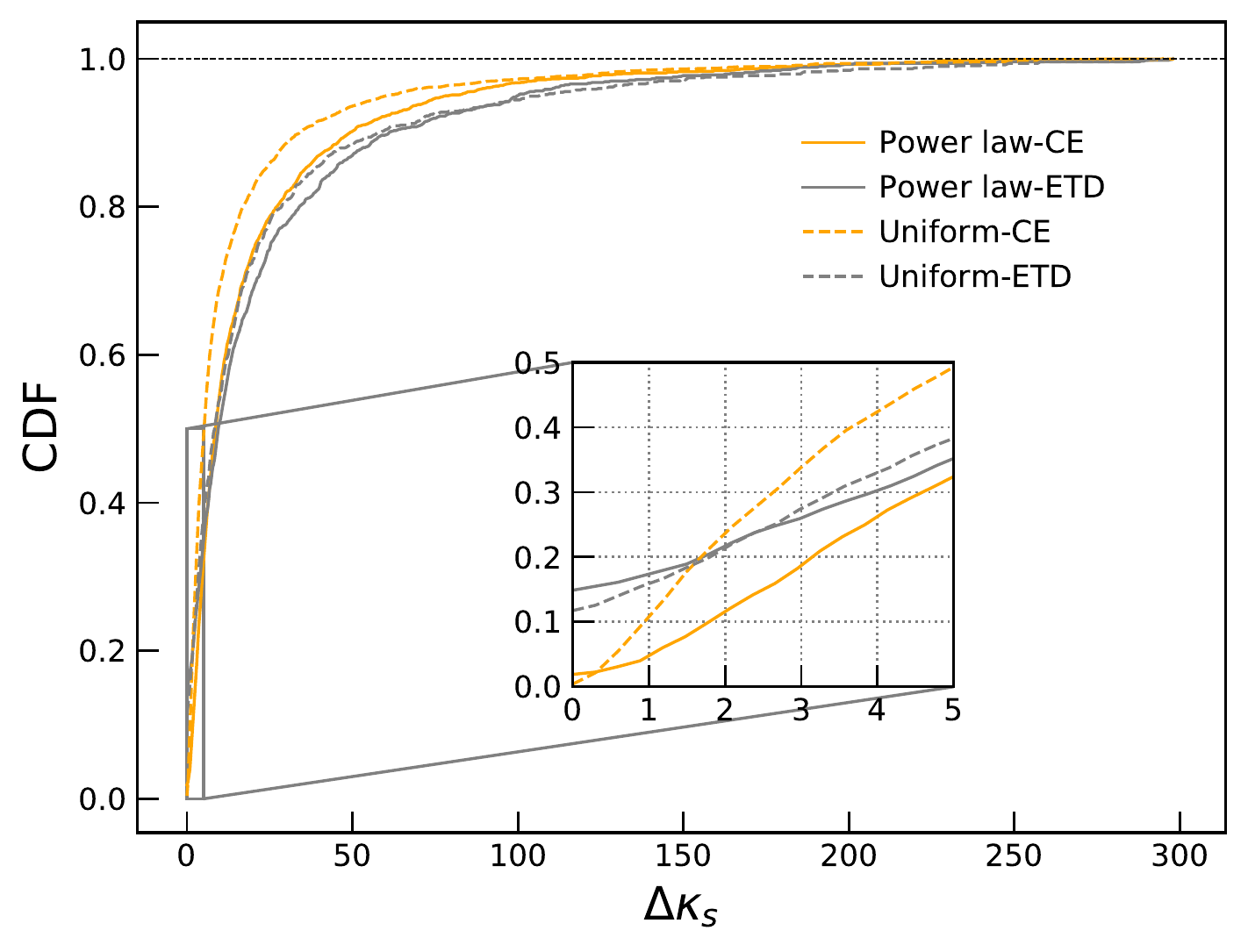}
\end{center} 
\caption{ The cumulative distribution function of errors on $\kappa_s$ for two
prototypical  astrophysical populations 
of binary black holes corresponding to two different models for the
binary's mass distribution. In the first model we
assume both component masses  to be uniformly 
distributed between $5M_{\odot}$ to $20M_{\odot}$ while the second model
assumes the primary mass to follow a power-law distribution with an
index $\alpha=2.3$~\cite{LIGOScientific:2018mvr,GW170104} and uniform distribution for the secondary. In both the
models the masses are defined with respect to the source frame and the  
sources are distributed uniformly  in the comoving volume up to a
redshift of 1.}
\label{Ks_population}
\end{figure}

\section{Parameter estimation using the Fisher Information Matrix  analysis}
\label{sec2:PE}
When we have an accurate model for the signal of interest and the expected
sensitivity of the detector, Fisher information matrix approach can be used to
compute the  $1$-$\sigma$ error bars on the parameters of the
signal~\cite{CutlerFlanagan1994} assuming the noise in the detector is
Gaussian-stationary and the signal-to-noise ratio is high. Here we employ this
approach to estimate the possible error bars on parameters associated with
spin-induced multipole moment of the compact binary system. A quick review of
Fisher information matrix formalism is given here. More details can be found in
\cite{CutlerFlanagan1994}.

A detector output consisting of the gravitational wave signal and
the background noise can be written as,
\begin{eqnarray} {s}(t)=h(t;{\theta_i})+n(t), \label{eqn3} \end{eqnarray}
where $h(t;{{\theta}_i})$ is the true signal which is buried in the noise
$n(t)$ and ${{\theta}_i}$ represents the set of parameters that characterizes
the signal. Due to the presence of noise, the measured parameters $\theta_i$
can fluctuate about the true value leading to errors associated with their
measurements. Hence measured value of $\theta_i=\theta_i^{\rm true}\pm \Delta
\theta_i$, where $\theta_i^{\rm true}$ is the true value of the parameter and
$\Delta \theta_i$ is the error associated with the measurement due to noise,
give us information about the parameter $\theta_{i}$. From the measurement, we
are interested in the probability distribution function for $\theta_i$ given
the signal $s(t)$, $p({{\theta}_i}|s)$. It can be shown that, for Gaussian
noise in the limit of high signal-to-noise ratios, the posterior probability
takes the form,
\begin{eqnarray} p({\theta_i}|s) \propto
\e^{-\frac{1}{2}(\Gamma_{jk}\Delta{\theta_{j}} \Delta{\theta_{k}})}, \label{eqn5}
\end{eqnarray}
where $\Gamma_{ij}$ is called the Fisher information matrix~\cite{Rao45,
Cramer46} defined as follows,
\begin{eqnarray} \Gamma_{ij}=2 \int^{f_{\rm upper}}_{f_{\rm lower}} df
\frac{{\tilde h_{i}}(f){\tilde h_{j}}^{\ast}(f)+ {\tilde h_{j}}(f){\tilde
h_{i}}^{\ast}(f)}{S_{n}(f)}, \label{eqn6} \end{eqnarray}
where $S_n(f)$ represents the noise power spectral density (PSD) of the detector and
$\tilde{h}_{i}\equiv{\partial\tilde{h}(f;{\theta_{i})}}/{\partial {\theta_i}}$
is evaluated at the true value of the parameter ${\theta}_{i}={\theta}_{i}^{\rm
true}$. Inverse of the Fisher information matrix is called the covariance
matrix ($\Sigma_{ij}$) and the error $(\sigma_{i})$ on each  parameter ${\theta_{i}}$ is given
by the square root of the diagonal entries of the covariance matrix. That is, 
\begin{eqnarray} \sigma_{i}=\sqrt{\Sigma_{ii}}. \label{eqn7} \end{eqnarray}

We choose to terminate the integral of Eq.~\eqref{eqn6} at twice the orbital
frequency of the inner most stable circular orbit ($f_{\rm ISCO}$) for a {\it
spinning} compact binary and use the fits obtained in
\cite{Husa:2015iqa,EccPEFavata}\footnote{Here we only consider contributions from the
second harmonic as discussed in Sec.~\ref{sec:waveformmodel}.}. The lower
frequency cut-off in the integral of Eq.~\eqref{eqn6} is fixed by the
sensitivity of the detector given by the function $S_n(f)$. In this work, we
intend to explore the parameter estimation analysis for two different
third-generation gravitational wave detector configurations: Cosmic Explorer (CE)
\cite{Evans:2016mbw, 0264-9381-28-9-094013}  and Einstein Telescope (ET-D)~\cite{0264-9381-28-9-094013}. Since the two have comparable sensitivities
and we choose one of them (in our case CE noise PSD) for the most
part of the paper. However, we compare the performance of CE and ET-D for a few
representative cases. The low frequency cut-off for CE (ET-D) configuration is
chosen to be  $5$Hz ($1$Hz) which defines the $f_{\rm low}$ value we use in the
integral given in Eq.~\eqref{eqn6}. We also discuss the improvements one expect
due to the use of third-generation detector sensitivities over advanced LIGO
and choose low frequency cut-off as  20 Hz for advanced LIGO.

\begin{figure}
\includegraphics[scale=0.55]{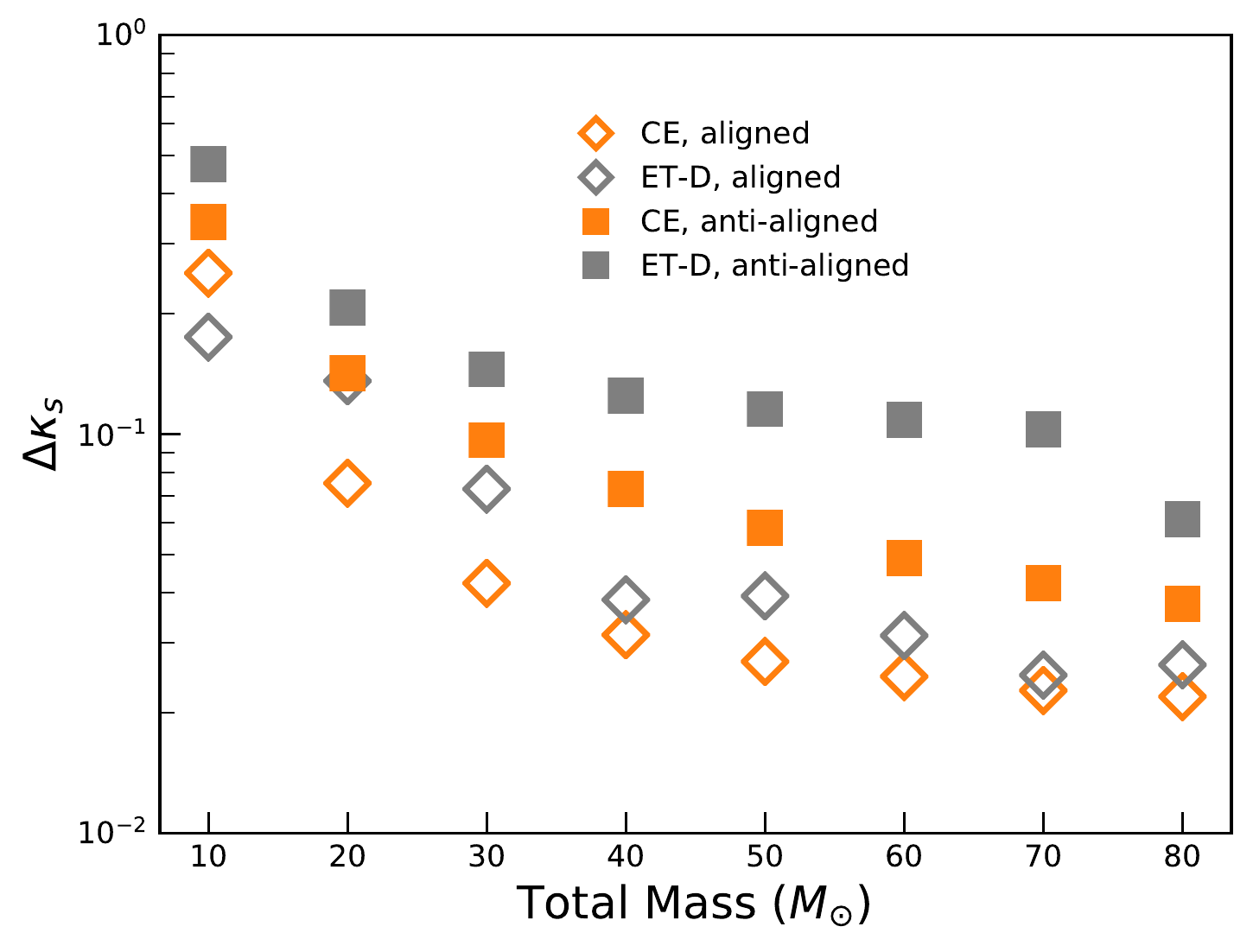}
\caption{Errors on the $\kappa_s$ as a function of the total mass of the binary
system for two representative 3rd generation detectors, Cosmic Explorer
(CE noise PSD) and Einstein Telescope (ET-D noise PSD). The binary is assumed to be at a distance of 400Mpc and is optimally oriented. The binary's mass-ratio is 1.2 and spin magnitudes of 0.9 and 0.8 for heavier and
lighter components, respectively. Filled- (empty-) markers represent
spin orientations of each component aligned (anti-aligned) to the orbital angular momentum while squares (diamonds) represent error
estimates for Cosmic Explorer (Einstein Telescope, ET-D).}
\label{Ks_DiffMassRatios_SpinOri_ET-D_Comp} \end{figure}
\begin{figure*}[hbtp]
\begin{center}
	\includegraphics[scale=0.3]{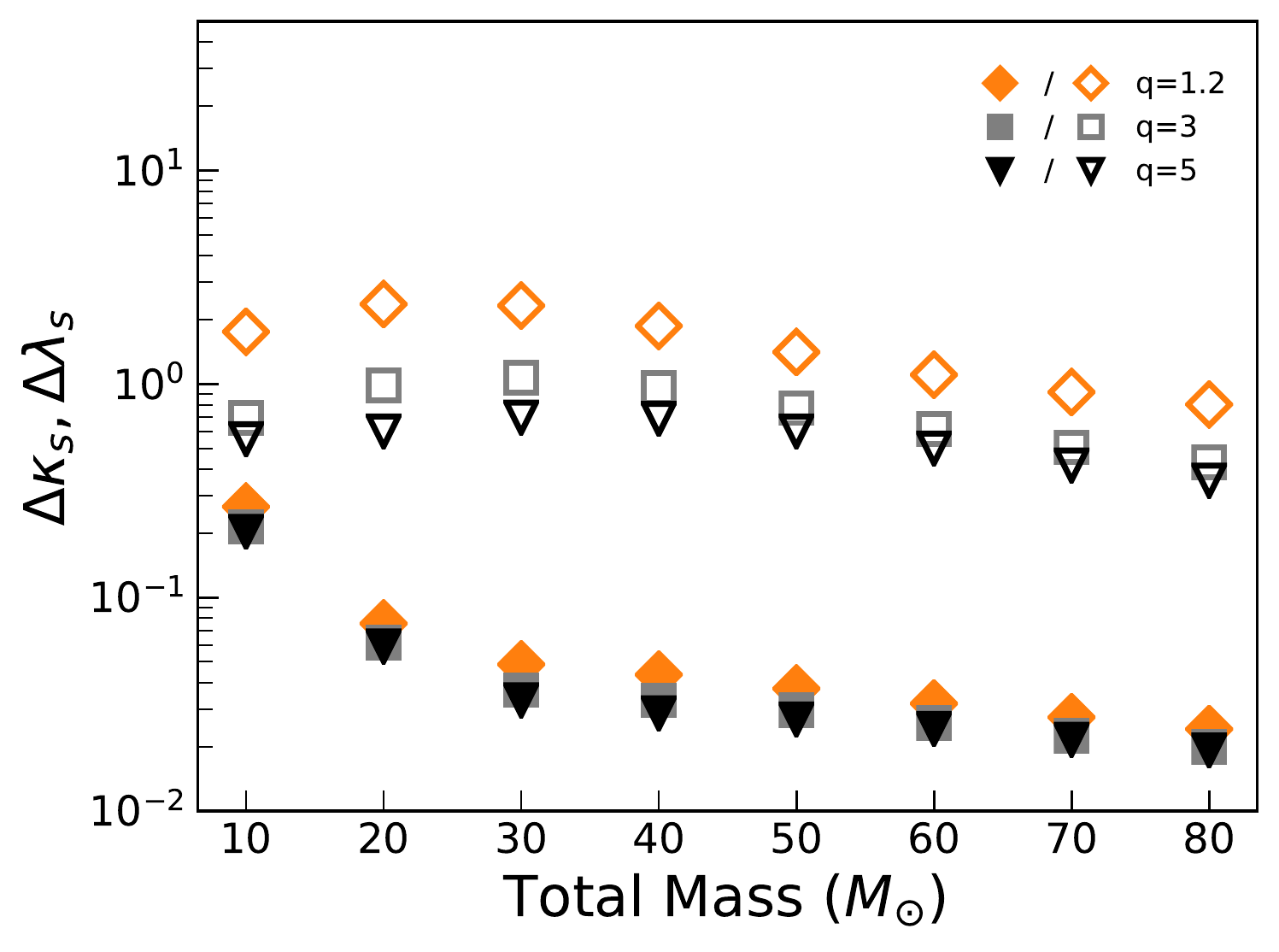}
	\includegraphics[scale=0.3]{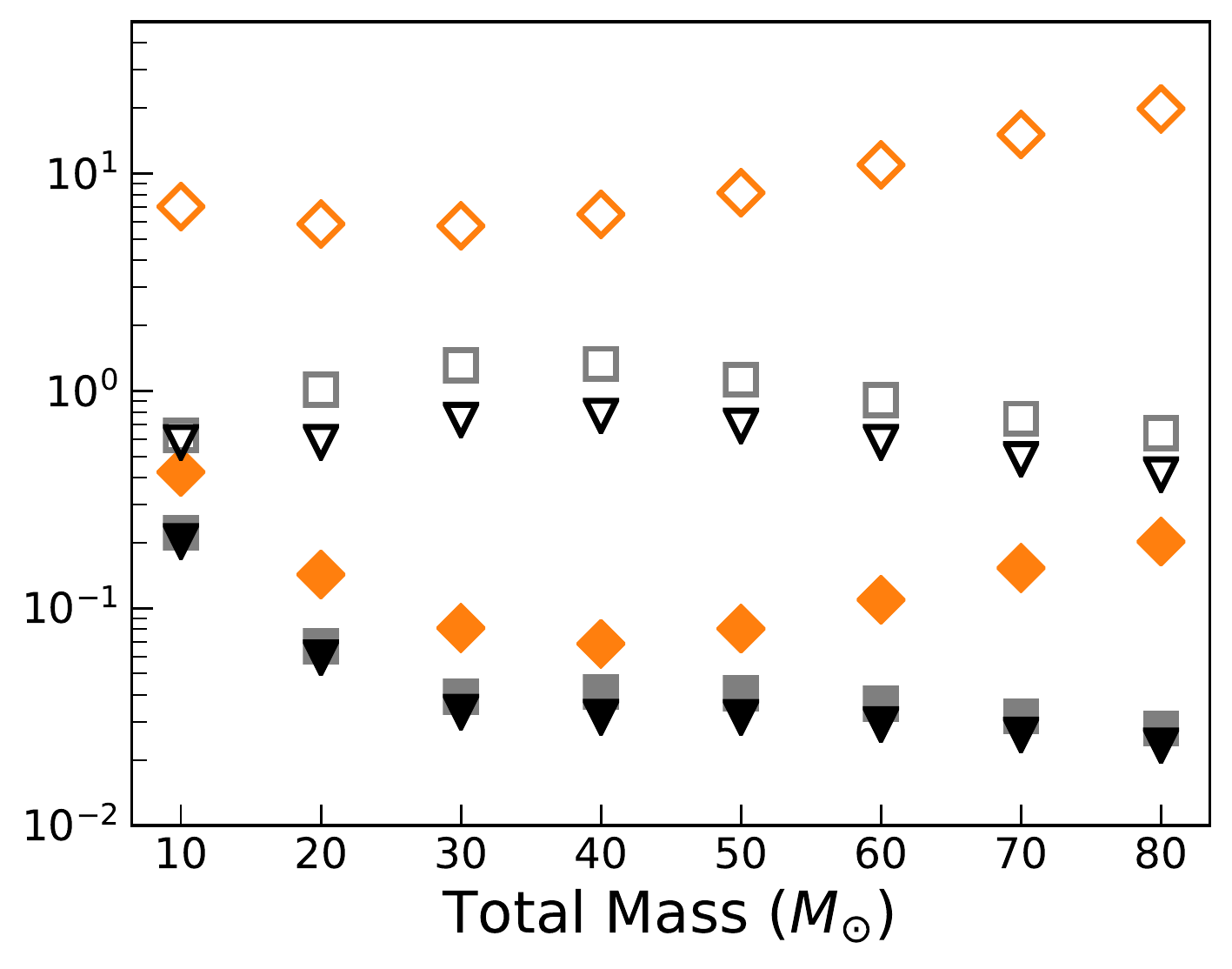}
	\includegraphics[scale=0.3]{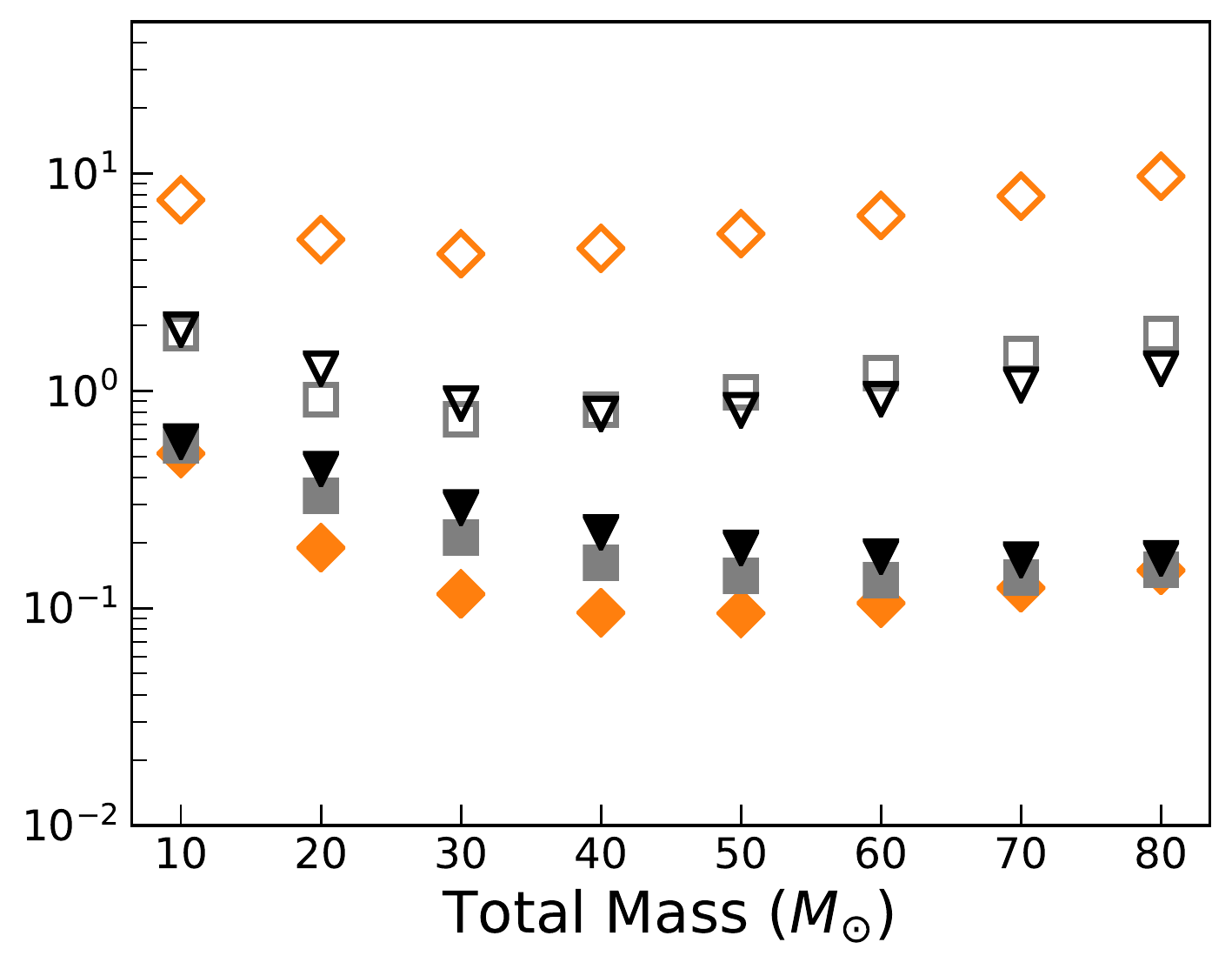}
	\includegraphics[scale=0.3]{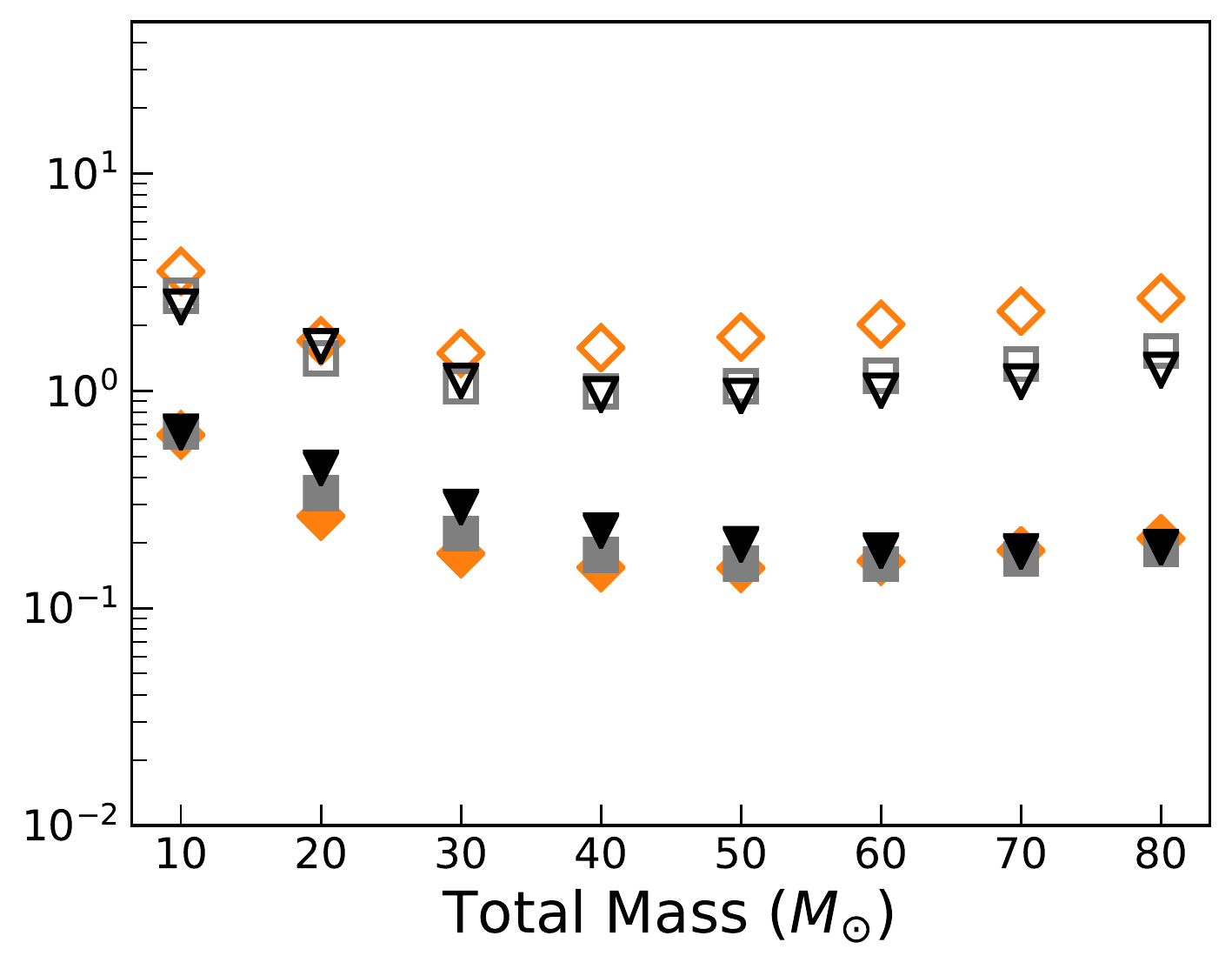}
\end{center} 
\caption{ Figure displays variation of $1-\sigma$ errors on
		$\kappa_s$ (filled markers) and $\lambda_s$ (unfilled markers) as a
		function of the binary's total mass for three representative mass-ratio
		cases and four representative spin-orientations with fixed component
		spin magnitudes ($\chi_1,\chi_2$) of $(0.9, 0.8)$. The four panels
		(left to right) represent binaries where spins of the two BHs are
		aligned, heavier one aligned and the other anti-aligned, heavier one
		anti-aligned and the other aligned and both the spins are anti-aligned
		to the orbital angular momentum axis. We assume the binary to be optimally oriented at a luminosity distance of 400Mpc. } \label{Ks_Ls_DiffMassRatios}
\end{figure*}
\section{Results and discussions}
\label{sec:results}

In this section, we present the results of our analyses. We perform
the parameter estimation analysis for a set of prototypical (stellar mass)
compact binary systems with the assumption that the binaries are optimally oriented and
are located at a {\it fiducial} distance of $400$ Mpc. The component spin
magnitudes are represented by the dimensionless spin parameter, $\chi_{1,\,2}$,
where subscripts 1(2) represents the primary (secondary) binary component. We also
follow the convention to assign higher mass and spin values to the primary
component. In addition to this, we also obtain a distribution
of errors of the spin-induced multipole moment parameters for a
simulated population of binary black holes, which act as proxies for the
binary black hole population third generation detectors would observe.

As discussed above, we choose to work with the  Cosmic Explorer noise
PSD as a representative noise sensitivity of a third-generation detector
configuration  \cite{Regimbau:2012ir, Hild:2010id,Hild:2008ng}. The lower
(upper) frequency cut-offs appearing in Eq.~\eqref{eqn6} are chosen to be 5Hz
($2\times f_{\rm ISCO}$ for {\it spinning} binary black holes \cite{Husa:2015iqa,EccPEFavata}). These results are compared
with the corresponding ones for advanced LIGO and Einstein Telescope for a
selected set of binary configurations.

\subsection{Bounds on binary's spin-induced quadrupole moment parameter}
\label{subsec:kappa_s}

If we assume the two objects in the binary system suffer equal deformation due
to their individual spins ( {\it{i.e}}., $\kappa_{1}=\kappa_2$), the symmetric
combination of the coefficient of spin-induced quadrupole moments,
$\kappa_{s}$, will be the suitable parameter to constrain the binary black hole nature~\citep{Krishnendu:2017shb}. Any deviation from the binary black hole value of $\kappa_s=1$
can be interpreted as a possible  constraint on the binary black hole nature of the compact
binary system. The parameter space considered here is the following,
\begin{eqnarray} {{\theta}_i}=\lbrace {\,t_c,\,\phi_c,
{M}_c,\,{\eta},\,{\chi_{1}},\,{\chi_{2}},\,{\kappa}_s}\rbrace, \label{eqn11a}
\end{eqnarray}
where $t_{c}$ and $\phi_{c}$ are the time and phase at coalescence, $M_{c}$
$(M_{c}=M\,\eta^{3/5})$ is the chirp mass,
$\eta=\frac{m_{1}\,m_{2}}{(m_{1}+m_{2})^2}$ is the symmetric mass-ratio,
$M=m_1+m_2$ is the total mass and $m_1,m_2$ and $\chi_{1}$, $ \chi_{2}$  are
the masses and  dimensionless spin parameters of the binary constituents. Note
that, here $\kappa_s$ is the only spin-induced parameter that is considered
{\it free} in the analysis; other combinations, ($\kappa_a$, $\lambda_s$,
$\lambda_a$), are set to their binary black hole values of $\kappa_a=0$, $\lambda_s=1$ and
$\lambda_a=0$.

Figure~\ref{Ks_DiffMassRatios_SpinOri} shows the variation of the errors in the
measurement of the parameter  $\kappa_s$, as a function of the total mass of
the binary. These errors also provide us 1$-\sigma$ upper bounds
on the value of $\kappa_s$. Three different set of markers in the {\it top} panel plot
correspond to three different mass-ratios (q=1.2, 3, 5) while the
component spins are fixed to the values of $\chi_1=0.9$, $\chi_2=0.8$. On the
other hand, the {\it bottom} panel assumes a binary with fixed mass-ratio
($q=1.2$) and displays the errors for four different spin configurations. Each
set of markers in both panels suggest that errors decrease as the binary's mass
increases. This is largely due to larger signal-to-noise ratios associated
with heavier binaries with fixed mass-ratio and component spins. In addition,
the trends displayed in the {\it top} panel suggest improved $\kappa_s$
estimates for larger mass-ratio cases (though the improvement is very minor)
while those in the {\it bottom} panel show that the best $\kappa_s$ estimates
correspond to the case when the two objects have component spins aligned to the
orbital angular momentum. The improved $\kappa_s$ estimates with respect to the mass 
ratio may be attributed to the larger number of gravitational wave cycles 
for asymmetric systems in the detector band. Similarly, as the upper cut-off frequency for aligned spin
configuration is larger, leading to larger number of gravitational wave
cycles, the error estimates for aligned spin configurations are the
best.

Figure~\ref{Ks_7by7_AdvLIGO_CE} explores $\kappa_s$ error estimates in
component spin parameter space for a binary with total mass of 30$M_\odot$ and
mass-ratios of 1.2 (top panel) and 3 (bottom panel). Solid (dashed) contours
represent errors on $\kappa_s$ in the context of CE (advanced LIGO) detector.  We can compare the performance of advanced LIGO and CE at those points
where their contours intersect. It is obvious from the figure that the typical
improvements in the estimation of $\kappa_s$ due to CE is by a factor of $\sim40-50$. This improvement is correlated with the increased signal-to-noise ratio of the sources in
the CE band compared to advanced LIGO. It is worth noting that, even though the overall
improvement in sensitivity of CE over advanced LIGO is roughly a factor of 10,
due to the larger band width of CE, the signal-to-noise ratios are
higher than advanced~LIGO roughly by $\sim40-50$ which explains
the overall improvement in the parameter estimation of CE with respect
to advanced~LIGO.
 
Another striking feature in Fig.~\ref{Ks_7by7_AdvLIGO_CE} is the shape
of the contours in the component spin plane. For nearly equal mass
systems ($q=1.2$), both advanced LIGO and CE contours are nearly circular,
whereas for $q=3$ they are ellipses. This feature may be
explained by a close inspection of the structure of the leading order
spin-spin dependence in the phasing which is proportional to the
$\kappa_s$ parameter. The term schematically reads as
$\Psi_{\rm{spin-spin}}$ $\sim$ $\kappa_{s}\,\zeta(\eta,\,\chi_{1},\,\chi_{2})$, where 
\begin{equation}
\zeta(\eta,\chi_1,\chi_2)=\alpha(\eta) \,\chi_{1}^2 +
\beta(\eta)\,\chi_{2}^2.
\end{equation}
Here
\begin{subequations}
\begin{equation} 
\alpha(\eta)=\left(1 + \sqrt{1 - 4\, \eta}-
2\, \eta\right)
\end{equation}
and
\begin{equation} 
\beta(\eta)=\left(1 - \sqrt{1 - 4\, \eta}-2\,
\eta\right).
\end{equation}
\label{eq:alpha-beta}
\end{subequations}

The derivative of the waveform with respect to $\kappa_s$ now will scale as $\sim \zeta$
and the corresponding Fisher information matrix element will scale as
$\Gamma_{\kappa_s\kappa_s}\sim\zeta^2$. Intuitively, as the error on
$\kappa_s$ is proportional to the square root of the inverse of the
Fisher matrix, we find $\Delta \kappa_s\sim \zeta^{-1}$.
Now the contours of constant errors in the component spin plane have the
form,
\begin{equation}
\chi_1^2
\left(\alpha\,\Delta\kappa_s\right)+\chi_2^2\left(\beta\,\Delta\kappa_s\right)=1.
\end{equation}
It is now obvious that for equal mass systems for which $\alpha=\beta$,
the contours of constant errors should be circles whereas for unequal mass systems the
contours will be ellipses. From Eq.~(\ref{eq:alpha-beta}), as $\frac{1}{\sqrt \alpha}\leq
\frac{1}{\sqrt{\beta}}$, these ellipses will have their semi-major axis
along $\chi_2$ direction as seen in the bottom panel of
Fig.~\ref{Ks_7by7_AdvLIGO_CE}. Though
this scaling completely neglects the correlation of  $\kappa_s$ with
other parameters, this does give us a qualitative picture about the
shape and orientation of the contours.

 	 We also explore the performance of the proposed test on an astrophysical population of binary black holes that the third-generation detectors may see by simulating two populations of binary black holes which correspond to different models for the component mass distribution. In the first model, we distribute the {\it source frame} component masses $m_{1,2}$ (here $m_{1}>m_{2}$) uniformly between 5$M_{\odot}$ and 20$M_{\odot}$. The second model assumes a power-law distribution with an index $\alpha=2.3$~\cite{LIGOScientific:2018mvr,GW170104} for the primary and uniform distribution for the secondary, again, with masses between 5$M_{\odot}$ to 20$M_{\odot}$. For both these cases, we distribute sources with constant comoving number density up to a redshift of $z=1$. The source locations and orientations are uniform on the sky and the polarization spheres, respectively. In order to account for the cosmological redshift on the gravitational signal we rescale the source frame masses ($m_s$) to redshifted masses ($m_d$) as, $m_{d}=m_{s}(1+z)$ in the gravitational wave signal while performing parameter estimation using Fisher matrix. This means that the maximum and minimum component masses in the detector frame will be $5M_{\odot}$ and $40M_{\odot}$, respectively. 
	We randomly draw 2000 sources from this population and perform 
the Fisher analysis to obtain the errors on various parameters including 
$\kappa_s$. Figure~\ref{Ks_population} shows the resulting distribution of errors on $\kappa_s$ for the two populations described above using Einstein Telescope and Cosmic Explorer.  As can be seen in the inset of Fig.~\ref{Ks_population}, use of the uniform over power-law 
distribution leads to nearly $20\%$ increase in the population of binaries observed with $\Delta \kappa_s\leq5$ for Cosmic Explorer whereas the errors we get using Einstein Telescope are largely independent of the mass distribution. Furthermore, we find that errors on $\kappa_s$ are less than 10 for $52\%$ ($68\%$) of the sources for the power-law (uniform) distribution model if we assume CE sensitivity. The numbers change to $41\%$ and $45\%$ respectively for power-law and uniform distributions when we consider Einstein Telescope. These trends can be understood as follows: 
the mass ratio distribution with primary's mass distributed using the power-law
 leads to {\it fewer} sources with larger mass ratios compared to the
case where we assume uniform distribution for component masses. In addition, 
the proposed test is more effective when the mass ratios are higher (see Fig.~\ref{Ks_DiffMassRatios_SpinOri}). 
These two factors improve the overall performance of the test for the uniform mass distribution as 
can be seen in  Fig.~\ref{Ks_population}.

Finally, Fig.~\ref{Ks_DiffMassRatios_SpinOri_ET-D_Comp} compares $\kappa_s$
estimates obtained using two different third-generation detector
configurations, Cosmic Explorer (CE) and Einstein Telescope (ET-D). In this
case, errors on $\kappa_s$ as a function of total mass for a fixed mass-ratio
of 1.2 is shown. We consider two spin orientations here, both the black holes
aligned and both the black holes anti-aligned to the orbital angular momentum
axis. As we expect,  the performance of CE and ET-D detectors are
comparable. However, the Cosmic Explorer error estimates are marginally better
than ET-D for all cases except at low masses when component spins are
aligned with respect to the orbital angular momentum. This
should be a reflection of the improved low frequency sensitivity of ET-D at
frequencies less than 5 Hz.

\subsection{Simultaneous bounds on binary's spin-induced quadrupole and octupole moment parameters}
\label{subsec:KL}

Below we discuss the measurability of both the quadrupolar and octupolar
spin-induced deformations due to individual BH spins, simultaneously. This time
we intend to measure a symmetric combination of coefficients characterizing the
spin-induced octupole moment of the compact binary system:
$\lambda_s=(\lambda_1+\lambda_2)/2$ along with the parameter $\kappa_s$. Again
the anti-symmetric combinations $\kappa_a$ and $\lambda_a$ are set to their binary black hole
value of zero. Formally, simultaneous bounds on $\kappa_{s}$ and $\lambda_{s}$
are more stringent than the $\kappa_{s}$ alone as we are sensitive to two of the leading spin-induced multipoles instead of one. The parameter space considered for this analysis is,
\begin{eqnarray} {{\theta}_i}=\lbrace
\,t_c,\,\phi_c,\,{M}_c,\,{\eta},\,{\chi_{1}},\,{\chi_{2}},\,{\kappa}_s,
\,{\lambda}_s\rbrace,  \label{eqn11c} \end{eqnarray}
where all the parameters have their usual meaning.

Figure~\ref{Ks_Ls_DiffMassRatios} shows variations in estimating bounds on
$\kappa_s$ (filled markers) and $\lambda_s$ (unfilled markers) as a function of
the total mass of the binary for three different mass-ratios ($q=1.2, 3, 5$)
and for fixed spin magnitudes of 0.9 and 0.8. Spin orientations chosen are
those where both the black hole spins aligned, heavier black hole spin aligned
and other anti-aligned, heavier black hole spin anti-aligned other aligned and
both the spins anti-aligned to the orbital angular momentum axis, respectively
from left to right of Fig.~\ref{Ks_Ls_DiffMassRatios}.

As discussed in Sec.~\ref{sec:waveformmodel}, spin-induced octupole
moment terms start to appear at 3.5 PN order in the PN phasing formula, while
the leading spin-induced quadrupole moment contributes at the 2PN order and
hence is a dominant effect in the PN dynamics. Hence, among $\kappa_s$ and
$\lambda_s$ the better constrained parameter is always $\kappa_s$.
From Fig.~\ref{Ks_Ls_DiffMassRatios}, it is clear that the $\kappa_s$ errors are
almost an order of magnitude better estimated compared to $\lambda_s$
errors and it is evident from the same figure that, for most of the parameter space, 
the errors on $\kappa_s$ is unaffected due to the inclusion of  $\lambda_s$ in the
problem.
 
 \begin{figure}[hbtp] 
\begin{center}
	\includegraphics[scale=0.5]{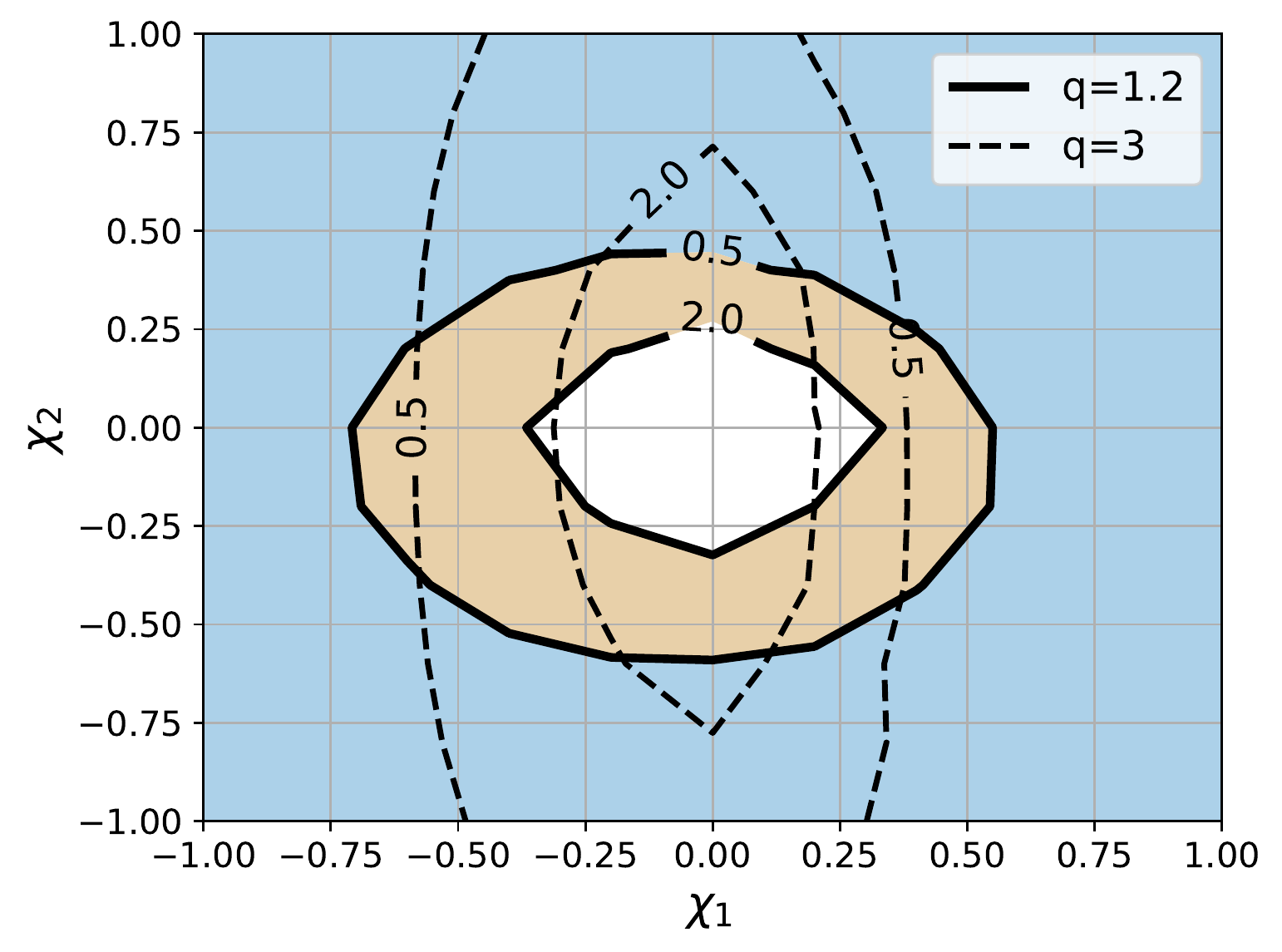}
	\includegraphics[scale=0.5]{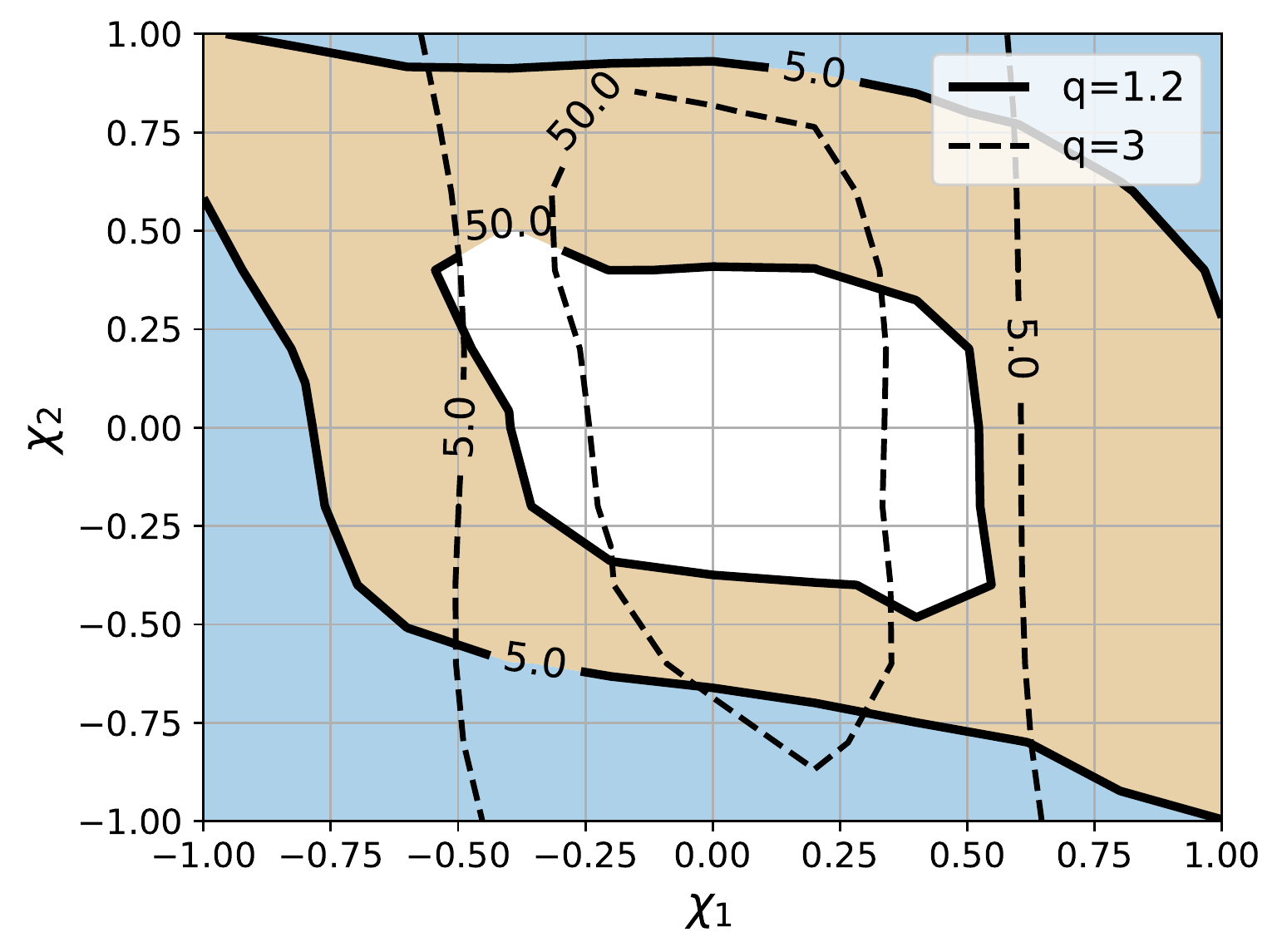} 
\end{center}
\caption{ Errors on spin-induced quadrupole and octupole
moment parameters of the binary-- $\kappa_s$ ({\it top panel}) and $\lambda_s$
({\it bottom panel}) in the $\chi_1$-$\chi_2$ plane for a binary system
with total mass $30M_{\odot}$. Solid contours represent mass-ratio of
1.2 and dashed ones represent mass-ratio 3. Binary system is assumed to be optimally oriented at a luminosity distance of 400Mpc.} \label{spincontourplot}
\end{figure}

\begin{figure*}[hbtp!]
	\includegraphics[scale=0.3]{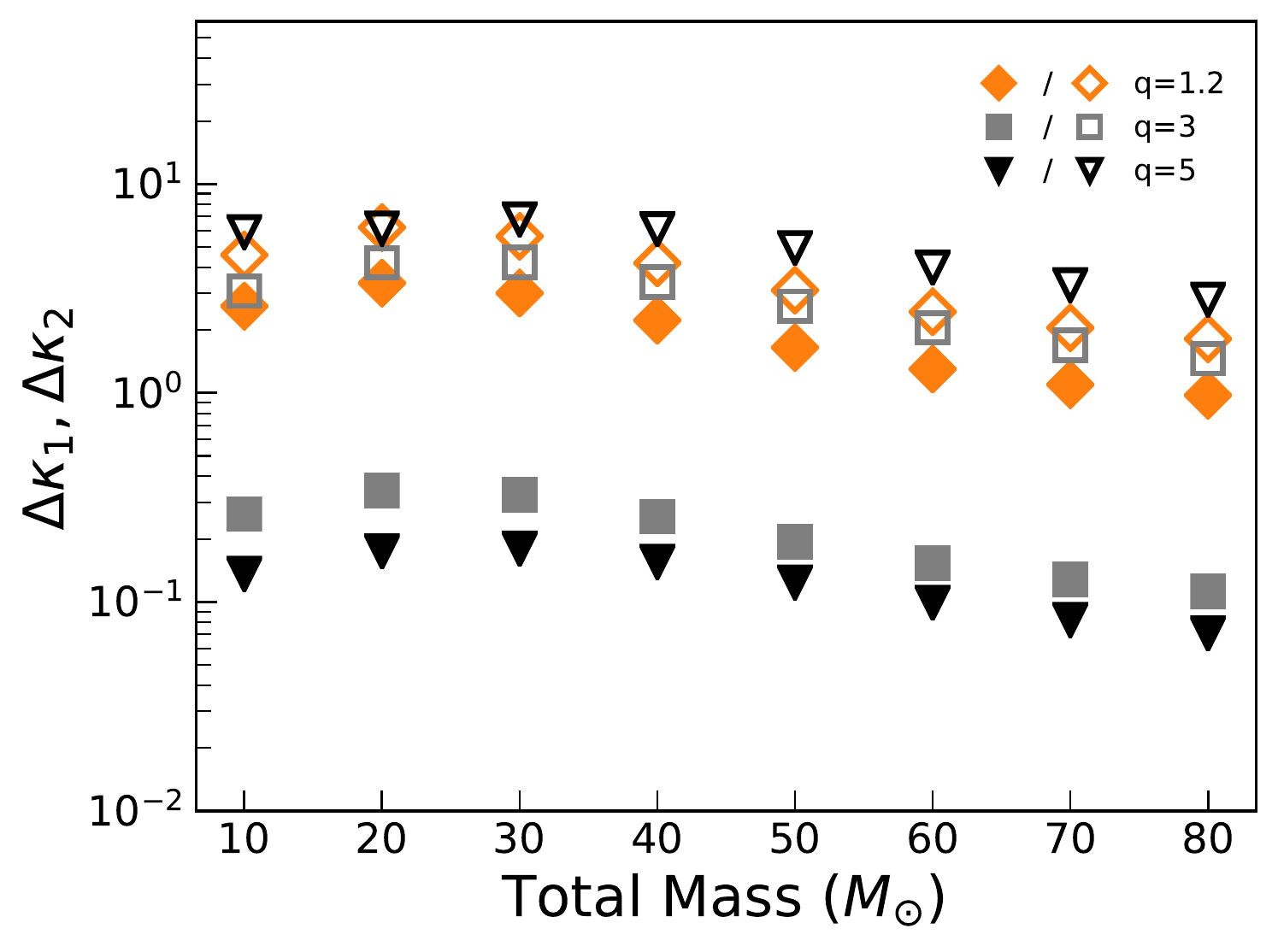}
	\includegraphics[scale=0.3]{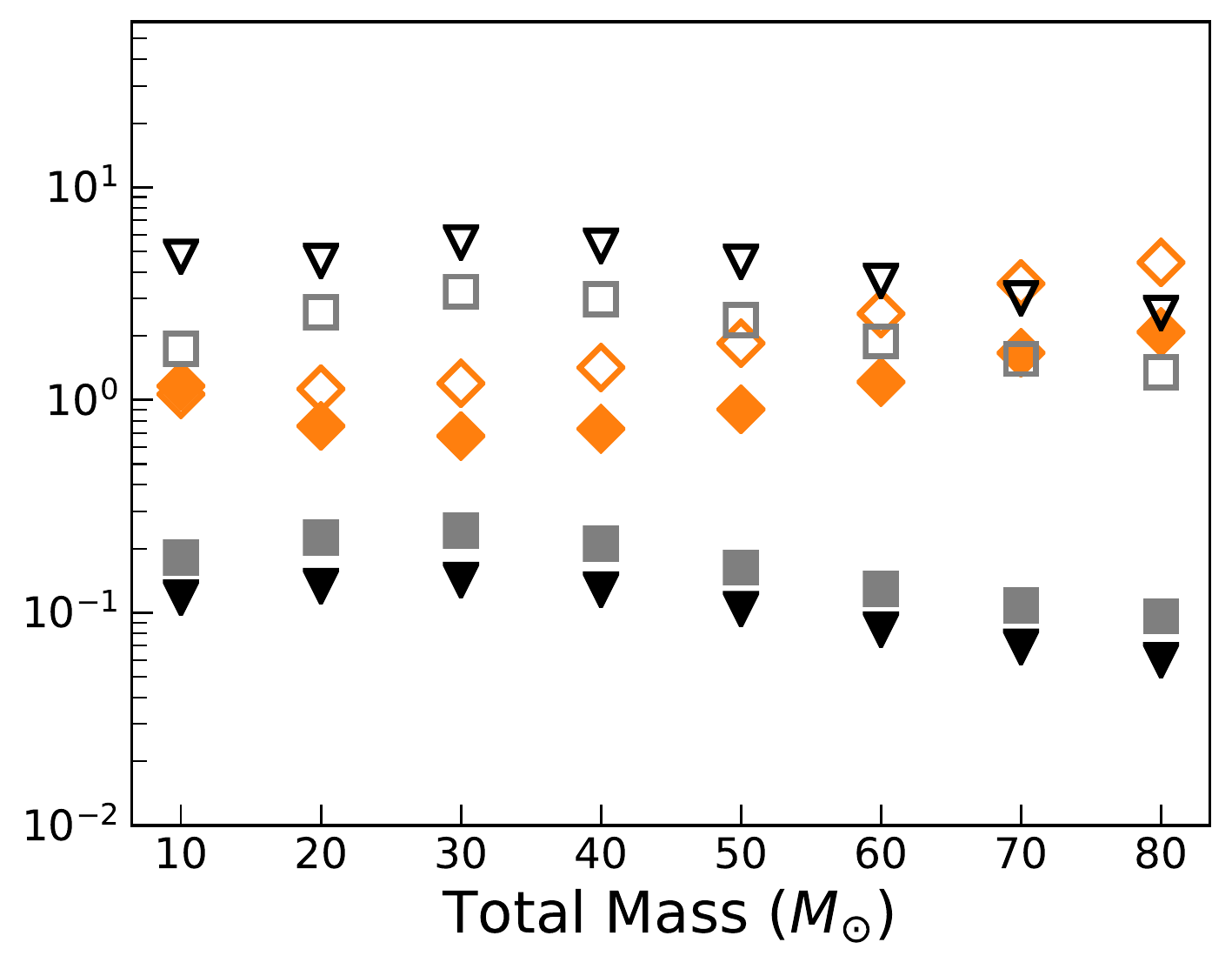}
	\includegraphics[scale=0.3]{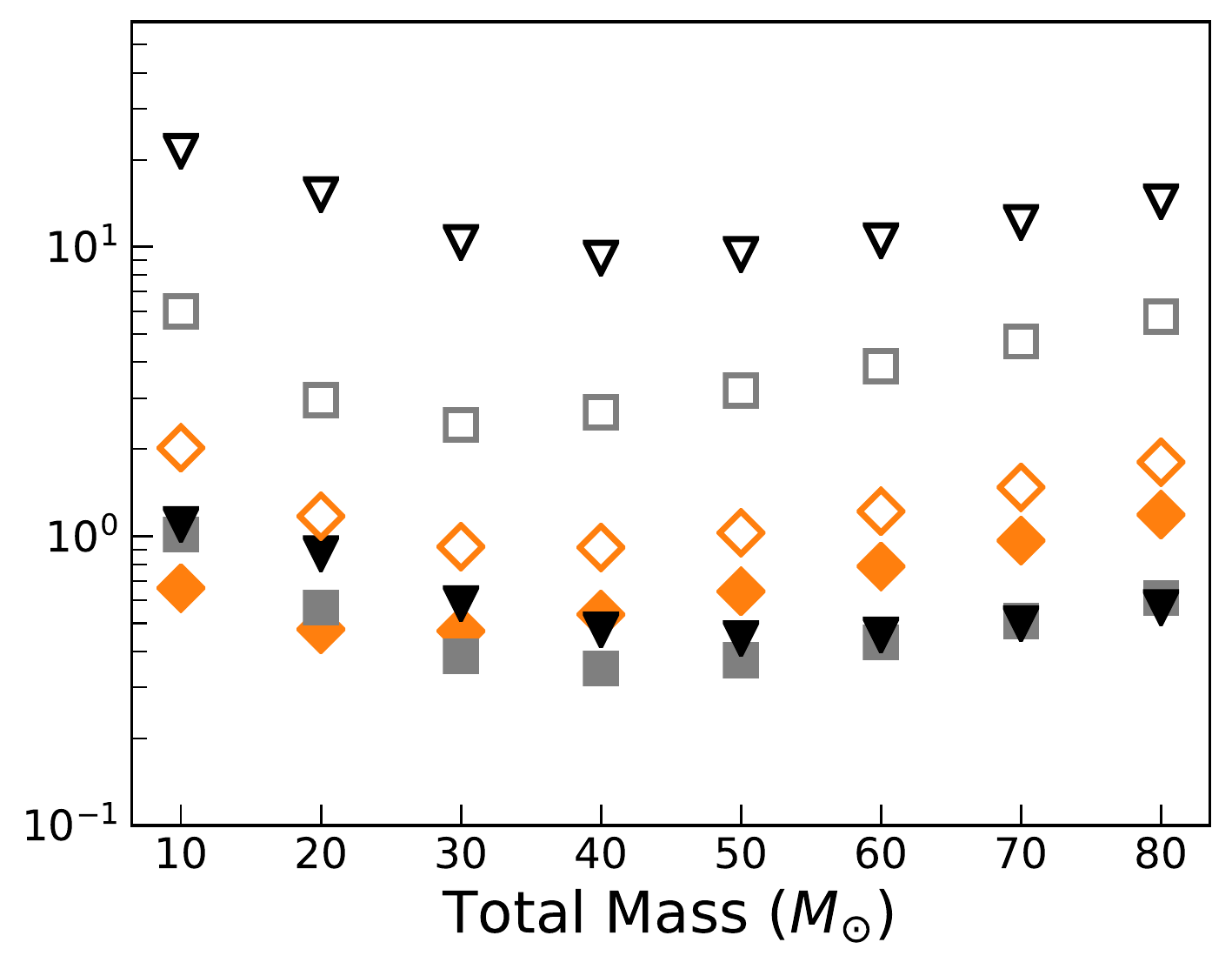}
	\includegraphics[scale=0.3]{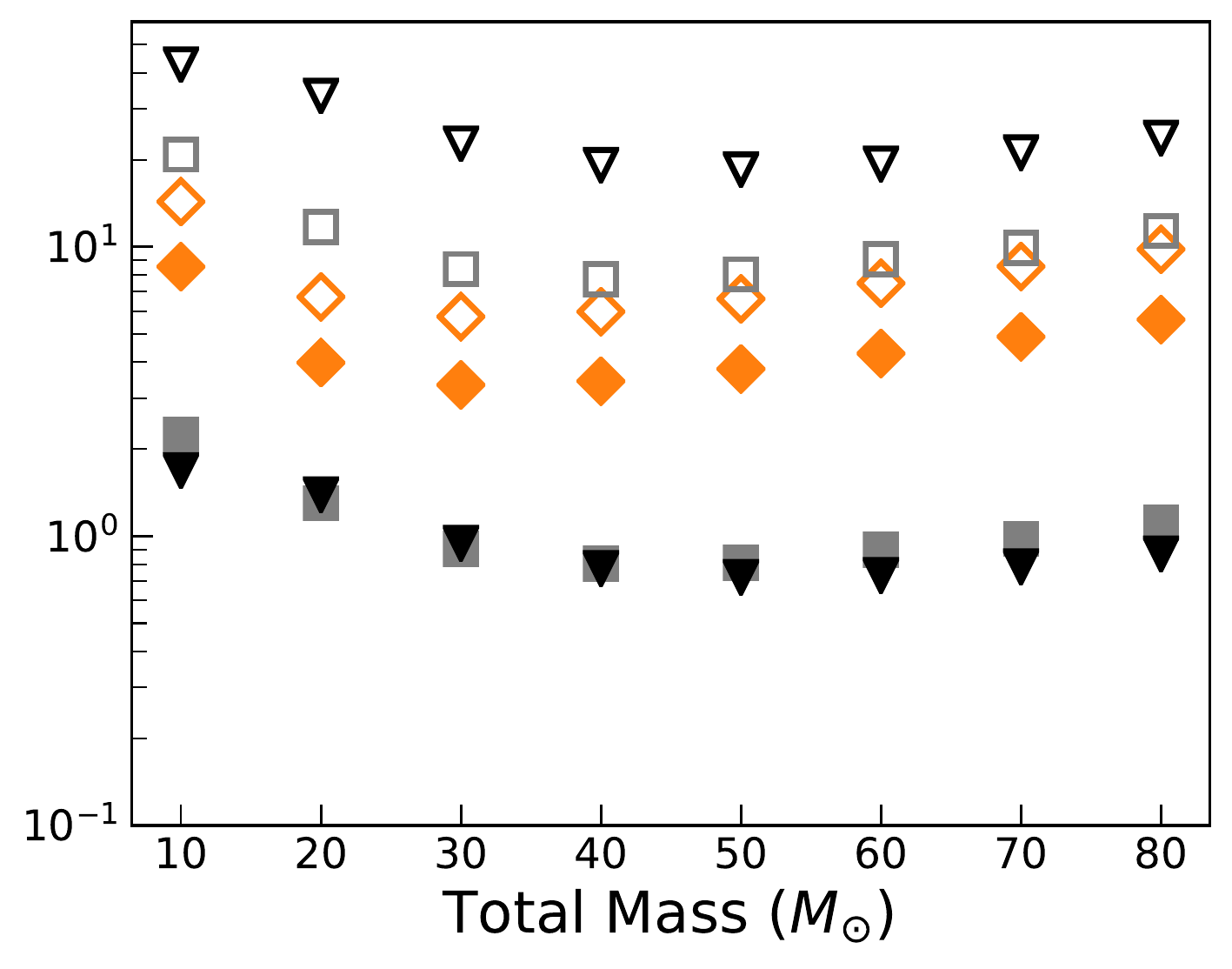}
\caption{Figure displays variation of $1-\sigma$ errors on $\kappa_1$
	(filled markers) and $\kappa_2$ (unfilled markers) as a
	function of the total mass of the binary system for {\it three}
	representative mass-ratio cases and {\it four} representative
	spin-orientations with fixed component spin magnitudes
	($\chi_1,\chi_2$) of $(0.9, 0.8)$. The {\it four} panels (left
	to right) represent binaries where spins of the two BH are
	aligned, heavier one aligned and the other anti-aligned,
	heavier one anti-aligned and the other aligned and both the
	spins are anti-aligned to the orbital angular momentum axis. 
	Also, note the (up-scaled) y-axes in last two panels.  } 
\label{fig:K1_K2}
\end{figure*}

Figure~\ref{Ks_Ls_DiffMassRatios} also shows that the bounds on both $\kappa_s$ and
$\lambda_s$ are tightly constrained for cases where the spin of the heavier
black hole aligned to the orbital angular momentum axis and if the binary is
more asymmetric. When both spins are aligned with respect to the orbital angular
momentum, the effect of mass-ratio is marginal (similar to the case presented in
Sec.~\ref{subsec:kappa_s} where only $\kappa_s$ is measured). On the other
hand, having the lighter component anti-aligned with respect to the orbital angular
momentum vector only marginally affects the measurements, with the most affected 
cases being the symmetric systems. We also note that the trends are not clear when we
deal with cases where heavier or both components are anti-aligned. In any case,
we do not expect the best results when heavier or both components are
anti-aligned.
 
The effect of spin magnitudes on the error estimates for simultaneous
$\kappa_s$ (top panel) and $\lambda_s$ (bottom panel) measurements are shown in
Fig.~\ref{spincontourplot}. We choose a total mass of $30M_{\odot}$ and mass
ratios of $q=1.2$ (solid contours) and $q=3$ (dotted contours) for this case. Broadly the
features seen here resemble those of Fig.~\ref{Ks_7by7_AdvLIGO_CE} where only
$\kappa_s$ was estimated. For nearly equal mass systems, we see that the contours are less
circular when $\lambda_s$ is included as a parameter. This may be due to
the degeneracies brought in by the estimation of $\lambda_s$. Regarding
the contours of constant error on $\lambda_s$ (bottom panel of
Fig.~\ref{spincontourplot}), following a line
of argument similar to the one in Sec.~\ref{subsec:kappa_s}, it can be
shown that the equations of the contours should schematically read as
$a\,\chi_1^3+b\,\chi_2^3=1$, where $a,\,b$ are functions of mass-ratio which decide
shape and orientation of the contours.

	 We performed an analysis, similar to the one reported in
Sec.~\ref{subsec:kappa_s}, where we simulated two populations of binary
black holes following a uniform and power-law distributions for the mass of
the binary's primary (heavier) component in the source frame, keeping the secondary
component mass to be uniformly distributed such that the total mass is
less than or equal to $40M_{\odot}$. We then compute the distribution of the bounds expected from the resulting population. Our analysis show that when $\kappa_s$ and $\lambda_s$ are measured simultaneously, errors on $\lambda_s$ are less than 10 for about $\sim6\%$($4\%$) sources when we use power-law (uniform) distribution on component masses for Cosmic Explorer. As observed earlier $\kappa_s$ estimates are marginally affected compared to the case when $\kappa_s$ alone is measured. We find that for nearly $42\% (51\%)$ sources $\Delta\kappa_s \leq10$ with power-law (uniform) distribution when measured along with $\lambda_s$. Again the error distribution for $\kappa_s$ is 
similar to those in Fig.~\ref{Ks_population}.

\subsection{Bounding the black hole nature of the compact binary constituents}
\label{sec:k1k2constraints}

In this section, we turn to our third and final analysis item -- measuring both
$\kappa_1$, $\kappa_2$ that characterize the spin-induced quadrupole moment coefficients of the binary components. Recall that simultaneous measurement of
both  $\kappa_1, \kappa_2$ provides a much stronger test compared to earlier
cases where we assumed the spin-induced multipole coefficients to be the same for
both the components of the binary ($\kappa_1=\kappa_2$, $\lambda_1=\lambda_2$).
The parameter space explored in this case is as follows,
 \begin{eqnarray} {{\theta}_i}=\lbrace { \,t_c,\,\phi_c,
 {M}_c,\,{\eta},\,{\chi_{1}},\,{\chi_{2}},\,{\kappa}_1, \,{\kappa}_2}\rbrace,
 \label{eqn11b} \end{eqnarray}
where the parameters have usual meaning. 

Figure~\ref{fig:K1_K2} shows
variations in errors on $\kappa_1$ (filled markers) and $\kappa_2$ (empty
markers) as a function of total mass of the binary for three different mass
ratios ($q=1.2, 3, 5$) and four different spin configurations (each with fixed
spin magnitudes of 0.9 and 0.8 for the heavier and lighter component,
respectively). Here again,  the spin orientations chosen are those where both
the black hole spins aligned, heavier component aligned and other anti-aligned,
heavier component anti-aligned other aligned and both the spins anti-aligned to
the orbital angular momentum axis, respectively from left to right of
Fig.~\ref{fig:K1_K2}. One of the first things we observe is that estimates of
$\kappa_1$ (which characterizes spin-induced deformations of the {\it heavier}
BH) is consistently better than those of  $\kappa_2$ (which characterizes
spin-induced deformations of the {\it lighter} BH) for all mass-ratios and spin
configurations. 
We also note that $\kappa_1$ is
measured with smaller errors for systems which are more asymmetric and if the
heavier BH is aligned with the orbital angular momentum axis.

These trends can be understood from the leading order spin-induced
quadrupole moment term in the gravitational wave phasing formula which
is proportional to
$\kappa_1\,\alpha(\eta)\,\chi_{1}^{2}+\kappa_2\,\beta(\eta)\,\chi_{2}^{2}$
as we explained in Sec.~\ref{subsec:kappa_s}. As $\alpha(\eta) \geq
\beta(\eta)$ and as we assign larger spin values to the more massive
component, for any given spin configuration the pre-factor of $\kappa_1$
is always higher than that of $\kappa_2$. This explains why $\kappa_1$
estimates are better than $\kappa_2$. Further, as $\alpha(\eta)$
increases with  mass-ratio, the error on $\kappa_1$ improves with mass
asymmetry. Similarly, the errors on $\kappa_2$ worsens with increase in the 
mass-ratio, since $\beta(\eta)$ is a decreasing function of the mass-ratio.

\section{Conclusion}
\label{sec:conclusion} 

In {\it three} different sets of numerical experiments discussed above,  we find
that improved sensitivities of third-generation detectors (Cosmic Explorer or
Einstein Telescope) over the current advanced LIGO detectors not only allow us
to significantly constrain the leading spin-induced effects in gravitational
waveforms but also enable us to explore a much wider mass and spin parameter
space (Sec.~\ref{subsec:kappa_s}). Assuming an astrophysical 
population of binary black holes, we show that the errors on spin-induced 
quadrupole moment parameter is $\leq$ 5 for  $30 \%$ of the 
total population if we assume the {\it primary} component masses follow the power-law distribution when CE configuration is used.
This fraction is roughly $20\%$ larger if the component masses are uniformly distributed while the errors with Einstein Telescope are largely independent of the 
mass distribution as can be seen from Fig.~\ref{Ks_population}.  
As expected, estimated bounds using the two
third-generation detectors (Cosmic Explorer or Einstein Telescope) are comparable with a
slight favor towards Cosmic Explorer configuration for high mass systems
whereas the low mass, aligned spins systems benefit the most from the improved
low frequency sensitivity of ET-D. We also showed that at least for a narrower
parameter space it would be possible to put stringent bounds on the first two
spin-induced multipole moments (quadrupolar and octupolar) simultaneously to
assess the nature of the involved compact binary (see Sec.~\ref{subsec:KL}
above). This also means one would be able to constrain {\it four} multipole
moments of the compact binary system facilitating a thorough probe of their binary black hole
nature. Finally, the possibility of bounding the leading spin-induced moment
for each binary component was explored in Sec.~\ref{sec:k1k2constraints}. Results of Fig.~\ref{fig:K1_K2} suggest a  set of possible binary 
configurations for which at least the nature of the heavier component 
can be confirmed.

Due to the dependencies of the bounds on component spins, mass
ratios, and masses, it is somewhat difficult to predict the magnitude of the
constraints these measurements will place on the parameter space of BH
mimickers. For example, considering boson star models of
Ryan~\cite{Ryan97b}, the theoretically allowed lower limit on the
quadrupole parameter is $\sim 10$ (see Fig. 4 of \cite{Ryan97b}) and
lower limit on the octupole parameter is $\sim 20$ (Fig. 5 of
\cite{Ryan97b}). These values do
depend on parameters such as the mass of the boson which constitute the
boson stars. Figure~\ref{Ks_7by7_AdvLIGO_CE} above shows that, unless component spin values are less than $0.1$, the expected bounds on $\kappa_s$ will rule out boson
star models which predict $\kappa_s\sim 10$. Similarly, as can be seen
from Fig.~\ref{spincontourplot}, detecting moderate to highly spinning
systems ($\chi\sim0.4-0.9$) can help you rule out $\lambda_s\sim20$.
However, it has to be borne in mind that the constraints from individual
events can rule out a BH mimicker association only for that system and
not a generic constrain on the parameter space of BH mimickers.
The constraints on gravastars, however, are likely to be weaker as
the spin-induced quadrupole moment parameter of this class of objects spans a small range
of values around the BH value ($\sim-0.17\textendash1.8$) as can be
seen from Fig. 7 of \cite{Uchikata:2015yma}.

{\it Acknowledgment:} The authors thank M. Saleem for useful discussions.
 We thank Anuradha Gupta for useful comments on the manuscript.
 K. G. A. and N. V. K. were partially supported by a grant from Infosys foundation.
K. G. A. acknowledges the Indo-US Science and Technology Forum through the
Indo-US {\em Centre for the Exploration of Extreme Gravity}, grant
IUSSTF/JC-029/2016. K. G. A. also acknowledges partial support by the Grant No.
EMR/2016/005594 of SERB.  This document has LIGO preprint number {\tt P1800325}. 
\bibliographystyle{apsrev}
\bibliography{references}

\begin{thebibliography}{97}
\expandafter\ifx\csname natexlab\endcsname\relax\def\natexlab#1{#1}\fi
\expandafter\ifx\csname bibnamefont\endcsname\relax
  \def\bibnamefont#1{#1}\fi
\expandafter\ifx\csname bibfnamefont\endcsname\relax
  \def\bibfnamefont#1{#1}\fi
\expandafter\ifx\csname citenamefont\endcsname\relax
  \def\citenamefont#1{#1}\fi
\expandafter\ifx\csname url\endcsname\relax
  \def\url#1{\texttt{#1}}\fi
\expandafter\ifx\csname urlprefix\endcsname\relax\def\urlprefix{URL }\fi
\providecommand{\bibinfo}[2]{#2}
\providecommand{\eprint}[2][]{\url{#2}}

\bibitem[{\citenamefont{Abbott et~al.}(2016{\natexlab{a}})}]{Discovery}
\bibinfo{author}{\bibfnamefont{B.~P.} \bibnamefont{Abbott}}
  \bibnamefont{et~al.} (\bibinfo{collaboration}{Virgo, LIGO Scientific}),
  \bibinfo{journal}{Phys. Rev. Lett.} \textbf{\bibinfo{volume}{116}},
  \bibinfo{pages}{061102} (\bibinfo{year}{2016}{\natexlab{a}}),
  \eprint{1602.03837}.

\bibitem[{\citenamefont{Abbott et~al.}(2016{\natexlab{b}})}]{GW151226}
\bibinfo{author}{\bibfnamefont{B.~P.} \bibnamefont{Abbott}}
  \bibnamefont{et~al.} (\bibinfo{collaboration}{Virgo, LIGO Scientific}),
  \bibinfo{journal}{Phys. Rev. Lett.} \textbf{\bibinfo{volume}{116}},
  \bibinfo{pages}{241103} (\bibinfo{year}{2016}{\natexlab{b}}),
  \eprint{1606.04855}.

\bibitem[{\citenamefont{Abbott et~al.}(2017{\natexlab{a}})}]{GW170104}
\bibinfo{author}{\bibfnamefont{B.~P.} \bibnamefont{Abbott}}
  \bibnamefont{et~al.} (\bibinfo{collaboration}{VIRGO, LIGO Scientific}),
  \bibinfo{journal}{Phys. Rev. Lett.} \textbf{\bibinfo{volume}{118}},
  \bibinfo{pages}{221101} (\bibinfo{year}{2017}{\natexlab{a}}),
  \bibinfo{note}{[Erratum: Phys. Rev. Lett.121,no.12,129901(2018)]},
  \eprint{1706.01812}.

\bibitem[{\citenamefont{Abbott et~al.}(2017{\natexlab{b}})}]{GW170608}
\bibinfo{author}{\bibfnamefont{B.~P.} \bibnamefont{Abbott}}
  \bibnamefont{et~al.} (\bibinfo{collaboration}{Virgo, LIGO Scientific}),
  \bibinfo{journal}{Astrophys. J.} \textbf{\bibinfo{volume}{851}},
  \bibinfo{pages}{L35} (\bibinfo{year}{2017}{\natexlab{b}}),
  \eprint{1711.05578}.

\bibitem[{\citenamefont{Abbott et~al.}(2017{\natexlab{c}})}]{GW170814}
\bibinfo{author}{\bibfnamefont{B.~P.} \bibnamefont{Abbott}}
  \bibnamefont{et~al.} (\bibinfo{collaboration}{Virgo, LIGO Scientific}),
  \bibinfo{journal}{Phys. Rev. Lett.} \textbf{\bibinfo{volume}{119}},
  \bibinfo{pages}{141101} (\bibinfo{year}{2017}{\natexlab{c}}),
  \eprint{1709.09660}.

\bibitem[{\citenamefont{Abbott
  et~al.}(2018{\natexlab{a}})}]{LIGOScientific:2018mvr}
\bibinfo{author}{\bibfnamefont{B.~P.} \bibnamefont{Abbott}}
  \bibnamefont{et~al.} (\bibinfo{collaboration}{LIGO Scientific, Virgo})
  (\bibinfo{year}{2018}{\natexlab{a}}), \eprint{1811.12907}.

\bibitem[{\citenamefont{Abbott
  et~al.}(2018{\natexlab{b}})}]{LIGOScientific:2018jsj}
\bibinfo{author}{\bibfnamefont{B.~P.} \bibnamefont{Abbott}}
  \bibnamefont{et~al.} (\bibinfo{collaboration}{LIGO Scientific, Virgo})
  (\bibinfo{year}{2018}{\natexlab{b}}), \eprint{1811.12940}.

\bibitem[{\citenamefont{Meidam et~al.}(2014)\citenamefont{Meidam, Agathos, Van
  Den~Broeck, Veitch, and Sathyaprakash}}]{Meidam:2014jpa}
\bibinfo{author}{\bibfnamefont{J.}~\bibnamefont{Meidam}},
  \bibinfo{author}{\bibfnamefont{M.}~\bibnamefont{Agathos}},
  \bibinfo{author}{\bibfnamefont{C.}~\bibnamefont{Van Den~Broeck}},
  \bibinfo{author}{\bibfnamefont{J.}~\bibnamefont{Veitch}}, \bibnamefont{and}
  \bibinfo{author}{\bibfnamefont{B.~S.} \bibnamefont{Sathyaprakash}},
  \bibinfo{journal}{Phys. Rev.} \textbf{\bibinfo{volume}{D90}},
  \bibinfo{pages}{064009} (\bibinfo{year}{2014}), \eprint{1406.3201}.

\bibitem[{\citenamefont{Yunes and Siemens}(2013)}]{Yunes:2013dva}
\bibinfo{author}{\bibfnamefont{N.}~\bibnamefont{Yunes}} \bibnamefont{and}
  \bibinfo{author}{\bibfnamefont{X.}~\bibnamefont{Siemens}},
  \bibinfo{journal}{Living Rev. Rel.} \textbf{\bibinfo{volume}{16}},
  \bibinfo{pages}{9} (\bibinfo{year}{2013}), \eprint{1304.3473}.

\bibitem[{\citenamefont{Arun et~al.}(2006{\natexlab{a}})\citenamefont{Arun,
  Iyer, Qusailah, and Sathyaprakash}}]{AIQS06a}
\bibinfo{author}{\bibfnamefont{K.~G.} \bibnamefont{Arun}},
  \bibinfo{author}{\bibfnamefont{B.~R.} \bibnamefont{Iyer}},
  \bibinfo{author}{\bibfnamefont{M.~S.~S.} \bibnamefont{Qusailah}},
  \bibnamefont{and} \bibinfo{author}{\bibfnamefont{B.~S.}
  \bibnamefont{Sathyaprakash}}, \bibinfo{journal}{Class. Quant. Grav.}
  \textbf{\bibinfo{volume}{23}}, \bibinfo{pages}{L37}
  (\bibinfo{year}{2006}{\natexlab{a}}), \eprint{gr-qc/0604018}.

\bibitem[{\citenamefont{Arun et~al.}(2006{\natexlab{b}})\citenamefont{Arun,
  Iyer, Qusailah, and Sathyaprakash}}]{AIQS06b}
\bibinfo{author}{\bibfnamefont{K.~G.} \bibnamefont{Arun}},
  \bibinfo{author}{\bibfnamefont{B.~R.} \bibnamefont{Iyer}},
  \bibinfo{author}{\bibfnamefont{M.~S.~S.} \bibnamefont{Qusailah}},
  \bibnamefont{and} \bibinfo{author}{\bibfnamefont{B.~S.}
  \bibnamefont{Sathyaprakash}}, \bibinfo{journal}{Phys. Rev.}
  \textbf{\bibinfo{volume}{D74}}, \bibinfo{pages}{024006}
  (\bibinfo{year}{2006}{\natexlab{b}}), \eprint{gr-qc/0604067}.

\bibitem[{\citenamefont{Berti et~al.}(2018{\natexlab{a}})\citenamefont{Berti,
  Yagi, and Yunes}}]{Berti:2018cxi}
\bibinfo{author}{\bibfnamefont{E.}~\bibnamefont{Berti}},
  \bibinfo{author}{\bibfnamefont{K.}~\bibnamefont{Yagi}}, \bibnamefont{and}
  \bibinfo{author}{\bibfnamefont{N.}~\bibnamefont{Yunes}},
  \bibinfo{journal}{Gen. Rel. Grav.} \textbf{\bibinfo{volume}{50}},
  \bibinfo{pages}{46} (\bibinfo{year}{2018}{\natexlab{a}}),
  \eprint{1801.03208}.

\bibitem[{\citenamefont{Berti et~al.}(2018{\natexlab{b}})\citenamefont{Berti,
  Yagi, Yang, and Yunes}}]{Berti:2018vdi}
\bibinfo{author}{\bibfnamefont{E.}~\bibnamefont{Berti}},
  \bibinfo{author}{\bibfnamefont{K.}~\bibnamefont{Yagi}},
  \bibinfo{author}{\bibfnamefont{H.}~\bibnamefont{Yang}}, \bibnamefont{and}
  \bibinfo{author}{\bibfnamefont{N.}~\bibnamefont{Yunes}},
  \bibinfo{journal}{Gen. Rel. Grav.} \textbf{\bibinfo{volume}{50}},
  \bibinfo{pages}{49} (\bibinfo{year}{2018}{\natexlab{b}}),
  \eprint{1801.03587}.

\bibitem[{\citenamefont{Will}(1977)}]{Will:1977wq}
\bibinfo{author}{\bibfnamefont{C.~M.} \bibnamefont{Will}},
  \bibinfo{journal}{Astrophys. J.} \textbf{\bibinfo{volume}{214}},
  \bibinfo{pages}{826} (\bibinfo{year}{1977}).

\bibitem[{\citenamefont{Agathos et~al.}(2014)\citenamefont{Agathos, Del~Pozzo,
  Li, Van Den~Broeck, Veitch, and Vitale}}]{TIGER2014}
\bibinfo{author}{\bibfnamefont{M.}~\bibnamefont{Agathos}},
  \bibinfo{author}{\bibfnamefont{W.}~\bibnamefont{Del~Pozzo}},
  \bibinfo{author}{\bibfnamefont{T.~G.~F.} \bibnamefont{Li}},
  \bibinfo{author}{\bibfnamefont{C.}~\bibnamefont{Van Den~Broeck}},
  \bibinfo{author}{\bibfnamefont{J.}~\bibnamefont{Veitch}}, \bibnamefont{and}
  \bibinfo{author}{\bibfnamefont{S.}~\bibnamefont{Vitale}},
  \bibinfo{journal}{Phys. Rev.} \textbf{\bibinfo{volume}{D89}},
  \bibinfo{pages}{082001} (\bibinfo{year}{2014}), \eprint{1311.0420}.

\bibitem[{\citenamefont{Yunes and Pretorius}(2009)}]{Yunes:2009ke}
\bibinfo{author}{\bibfnamefont{N.}~\bibnamefont{Yunes}} \bibnamefont{and}
  \bibinfo{author}{\bibfnamefont{F.}~\bibnamefont{Pretorius}},
  \bibinfo{journal}{Phys. Rev.} \textbf{\bibinfo{volume}{D80}},
  \bibinfo{pages}{122003} (\bibinfo{year}{2009}), \eprint{0909.3328}.

\bibitem[{\citenamefont{Will}(1998)}]{Will:1997bb}
\bibinfo{author}{\bibfnamefont{C.~M.} \bibnamefont{Will}},
  \bibinfo{journal}{Phys. Rev.} \textbf{\bibinfo{volume}{D57}},
  \bibinfo{pages}{2061} (\bibinfo{year}{1998}), \eprint{gr-qc/9709011}.

\bibitem[{\citenamefont{Samajdar and Arun}(2017)}]{Samajdar:2017mka}
\bibinfo{author}{\bibfnamefont{A.}~\bibnamefont{Samajdar}} \bibnamefont{and}
  \bibinfo{author}{\bibfnamefont{K.~G.} \bibnamefont{Arun}},
  \bibinfo{journal}{Phys. Rev.} \textbf{\bibinfo{volume}{D96}},
  \bibinfo{pages}{104027} (\bibinfo{year}{2017}), \eprint{1708.00671}.

\bibitem[{\citenamefont{Ghosh et~al.}(2017)\citenamefont{Ghosh,
  Johnson-Mcdaniel, Ghosh, Mishra, Ajith, Del~Pozzo, Berry, Nielsen, and
  London}}]{Ghosh:2017gfp}
\bibinfo{author}{\bibfnamefont{A.}~\bibnamefont{Ghosh}},
  \bibinfo{author}{\bibfnamefont{N.~K.} \bibnamefont{Johnson-Mcdaniel}},
  \bibinfo{author}{\bibfnamefont{A.}~\bibnamefont{Ghosh}},
  \bibinfo{author}{\bibfnamefont{C.~K.} \bibnamefont{Mishra}},
  \bibinfo{author}{\bibfnamefont{P.}~\bibnamefont{Ajith}},
  \bibinfo{author}{\bibfnamefont{W.}~\bibnamefont{Del~Pozzo}},
  \bibinfo{author}{\bibfnamefont{C.~P.~L.} \bibnamefont{Berry}},
  \bibinfo{author}{\bibfnamefont{A.~B.} \bibnamefont{Nielsen}},
  \bibnamefont{and} \bibinfo{author}{\bibfnamefont{L.}~\bibnamefont{London}}
  (\bibinfo{year}{2017}), \eprint{1704.06784}.

\bibitem[{\citenamefont{Abbott et~al.}(2016{\natexlab{c}})}]{TOG}
\bibinfo{author}{\bibfnamefont{B.~P.} \bibnamefont{Abbott}}
  \bibnamefont{et~al.} (\bibinfo{collaboration}{Virgo, LIGO Scientific}),
  \bibinfo{journal}{Phys. Rev. Lett.} \textbf{\bibinfo{volume}{116}},
  \bibinfo{pages}{221101} (\bibinfo{year}{2016}{\natexlab{c}}),
  \eprint{1602.03841}.

\bibitem[{\citenamefont{Abbott et~al.}(2016{\natexlab{d}})}]{O1BBH}
\bibinfo{author}{\bibfnamefont{B.~P.} \bibnamefont{Abbott}}
  \bibnamefont{et~al.} (\bibinfo{collaboration}{Virgo, LIGO Scientific}),
  \bibinfo{journal}{Phys. Rev.} \textbf{\bibinfo{volume}{X6}},
  \bibinfo{pages}{041015} (\bibinfo{year}{2016}{\natexlab{d}}),
  \eprint{1606.04856}.

\bibitem[{\citenamefont{{Blanchet}}(2014)}]{Blanchet}
\bibinfo{author}{\bibfnamefont{L.}~\bibnamefont{{Blanchet}}},
  \bibinfo{journal}{Living Reviews in Relativity}
  \textbf{\bibinfo{volume}{17}}, \bibinfo{eid}{2} (\bibinfo{year}{2014}),
  \eprint{1310.1528}.

\bibitem[{\citenamefont{Pretorius}(2007)}]{Pretorius:2007nq}
\bibinfo{author}{\bibfnamefont{F.}~\bibnamefont{Pretorius}}
  (\bibinfo{year}{2007}), \eprint{0710.1338}.

\bibitem[{\citenamefont{Sasaki and Tagoshi}(2003)}]{Sasaki:2003xr}
\bibinfo{author}{\bibfnamefont{M.}~\bibnamefont{Sasaki}} \bibnamefont{and}
  \bibinfo{author}{\bibfnamefont{H.}~\bibnamefont{Tagoshi}},
  \bibinfo{journal}{Living Rev. Rel.} \textbf{\bibinfo{volume}{6}},
  \bibinfo{pages}{6} (\bibinfo{year}{2003}), \eprint{gr-qc/0306120}.

\bibitem[{\citenamefont{{Giudice} et~al.}(2016)\citenamefont{{Giudice},
  {McCullough}, and {Urbano}}}]{Giudice}
\bibinfo{author}{\bibfnamefont{G.~F.} \bibnamefont{{Giudice}}},
  \bibinfo{author}{\bibfnamefont{M.}~\bibnamefont{{McCullough}}},
  \bibnamefont{and} \bibinfo{author}{\bibfnamefont{A.}~\bibnamefont{{Urbano}}},
  \bibinfo{journal}{jcap} \textbf{\bibinfo{volume}{10}}, \bibinfo{eid}{001}
  (\bibinfo{year}{2016}), \eprint{1605.01209}.

\bibitem[{\citenamefont{Sathyaprakash et~al.}(2011)}]{Sathyaprakash:2011bh}
\bibinfo{author}{\bibfnamefont{B.}~\bibnamefont{Sathyaprakash}}
  \bibnamefont{et~al.}, in \emph{\bibinfo{booktitle}{{Proceedings, 46th
  Rencontres de Moriond on Gravitational Waves and Experimental Gravity: La
  Thuile, Italy, March 20-27, 2011}}} (\bibinfo{year}{2011}), pp.
  \bibinfo{pages}{127--136}, \eprint{1108.1423}.

\bibitem[{\citenamefont{Regimbau et~al.}(2012)}]{Regimbau:2012ir}
\bibinfo{author}{\bibfnamefont{T.}~\bibnamefont{Regimbau}}
  \bibnamefont{et~al.}, \bibinfo{journal}{Phys. Rev.}
  \textbf{\bibinfo{volume}{D86}}, \bibinfo{pages}{122001}
  (\bibinfo{year}{2012}), \eprint{1201.3563}.

\bibitem[{\citenamefont{Hild et~al.}(2011{\natexlab{a}})}]{Hild:2010id}
\bibinfo{author}{\bibfnamefont{S.}~\bibnamefont{Hild}} \bibnamefont{et~al.},
  \bibinfo{journal}{Class. Quant. Grav.} \textbf{\bibinfo{volume}{28}},
  \bibinfo{pages}{094013} (\bibinfo{year}{2011}{\natexlab{a}}),
  \eprint{1012.0908}.

\bibitem[{\citenamefont{Hild et~al.}(2008)\citenamefont{Hild, Chelkowski, and
  Freise}}]{Hild:2008ng}
\bibinfo{author}{\bibfnamefont{S.}~\bibnamefont{Hild}},
  \bibinfo{author}{\bibfnamefont{S.}~\bibnamefont{Chelkowski}},
  \bibnamefont{and} \bibinfo{author}{\bibfnamefont{A.}~\bibnamefont{Freise}}
  (\bibinfo{year}{2008}), \eprint{0810.0604}.

\bibitem[{\citenamefont{{Mazur} and {Mottola}}(2004)}]{gravastar_Mazur}
\bibinfo{author}{\bibfnamefont{P.~O.} \bibnamefont{{Mazur}}} \bibnamefont{and}
  \bibinfo{author}{\bibfnamefont{E.}~\bibnamefont{{Mottola}}},
  \bibinfo{journal}{Proceedings of the National Academy of Science}
  \textbf{\bibinfo{volume}{101}}, \bibinfo{pages}{9545} (\bibinfo{year}{2004}),
  \eprint{gr-qc/0407075}.

\bibitem[{\citenamefont{{Liebling} and
  {Palenzuela}}(2012)}]{Bosonstar_Liebling}
\bibinfo{author}{\bibfnamefont{S.~L.} \bibnamefont{{Liebling}}}
  \bibnamefont{and}
  \bibinfo{author}{\bibfnamefont{C.}~\bibnamefont{{Palenzuela}}},
  \bibinfo{journal}{Living Reviews in Relativity}
  \textbf{\bibinfo{volume}{15}}, \bibinfo{eid}{6} (\bibinfo{year}{2012}),
  \eprint{1202.5809}.

\bibitem[{\citenamefont{Almheiri et~al.}(2013)\citenamefont{Almheiri, Marolf,
  Polchinski, and Sully}}]{Firewall}
\bibinfo{author}{\bibfnamefont{A.}~\bibnamefont{Almheiri}},
  \bibinfo{author}{\bibfnamefont{D.}~\bibnamefont{Marolf}},
  \bibinfo{author}{\bibfnamefont{J.}~\bibnamefont{Polchinski}},
  \bibnamefont{and} \bibinfo{author}{\bibfnamefont{J.}~\bibnamefont{Sully}},
  \bibinfo{journal}{JHEP} \textbf{\bibinfo{volume}{02}}, \bibinfo{pages}{062}
  (\bibinfo{year}{2013}), \eprint{1207.3123}.

\bibitem[{\citenamefont{{Hansen}}(1974)}]{Hansen74}
\bibinfo{author}{\bibfnamefont{R.~O.} \bibnamefont{{Hansen}}},
  \bibinfo{journal}{Journal of Mathematical Physics}
  \textbf{\bibinfo{volume}{15}}, \bibinfo{pages}{46} (\bibinfo{year}{1974}).

\bibitem[{\citenamefont{{Carter}}(1971)}]{Carter71}
\bibinfo{author}{\bibfnamefont{B.}~\bibnamefont{{Carter}}},
  \bibinfo{journal}{Physical Review Letters} \textbf{\bibinfo{volume}{26}},
  \bibinfo{pages}{331} (\bibinfo{year}{1971}).

\bibitem[{\citenamefont{Gürlebeck}(2015)}]{Gurlebeck:2015xpa}
\bibinfo{author}{\bibfnamefont{N.}~\bibnamefont{Gürlebeck}},
  \bibinfo{journal}{Phys. Rev. Lett.} \textbf{\bibinfo{volume}{114}},
  \bibinfo{pages}{151102} (\bibinfo{year}{2015}), \eprint{1503.03240}.

\bibitem[{\citenamefont{Ryan}(1995)}]{Ryan:1995wh}
\bibinfo{author}{\bibfnamefont{F.~D.} \bibnamefont{Ryan}},
  \bibinfo{journal}{Phys. Rev.} \textbf{\bibinfo{volume}{D52}},
  \bibinfo{pages}{5707} (\bibinfo{year}{1995}).

\bibitem[{\citenamefont{Geroch}(1970{\natexlab{a}})}]{Geroch2}
\bibinfo{author}{\bibfnamefont{R.}~\bibnamefont{Geroch}},
  \bibinfo{journal}{Journal of Mathematical Physics}
  \textbf{\bibinfo{volume}{11}}, \bibinfo{pages}{2580}
  (\bibinfo{year}{1970}{\natexlab{a}}).

\bibitem[{\citenamefont{Geroch}(1970{\natexlab{b}})}]{Geroch1}
\bibinfo{author}{\bibfnamefont{R.}~\bibnamefont{Geroch}},
  \bibinfo{journal}{Journal of Mathematical Physics}
  \textbf{\bibinfo{volume}{11}}, \bibinfo{pages}{1955}
  (\bibinfo{year}{1970}{\natexlab{b}}).

\bibitem[{\citenamefont{Binnington and
  Poisson}(2009)}]{RelativistictheoryofTLNs_EricBinnington}
\bibinfo{author}{\bibfnamefont{T.}~\bibnamefont{Binnington}} \bibnamefont{and}
  \bibinfo{author}{\bibfnamefont{E.}~\bibnamefont{Poisson}},
  \bibinfo{journal}{Phys. Rev. D} \textbf{\bibinfo{volume}{80}},
  \bibinfo{pages}{084018} (\bibinfo{year}{2009}).

\bibitem[{\citenamefont{{G{\"u}rlebeck}}(2015)}]{NHTBHsAstrophysicalEnvironments2015}
\bibinfo{author}{\bibfnamefont{N.}~\bibnamefont{{G{\"u}rlebeck}}},
  \bibinfo{journal}{Physical Review Letters} \textbf{\bibinfo{volume}{114}},
  \bibinfo{eid}{151102} (\bibinfo{year}{2015}), \eprint{1503.03240}.

\bibitem[{\citenamefont{{Flanagan} and
  {Hinderer}}(2008)}]{FlanaganHindererNSTLNs}
\bibinfo{author}{\bibfnamefont{{\'E}.~{\'E}.} \bibnamefont{{Flanagan}}}
  \bibnamefont{and}
  \bibinfo{author}{\bibfnamefont{T.}~\bibnamefont{{Hinderer}}},
  \bibinfo{journal}{\prd} \textbf{\bibinfo{volume}{77}}, \bibinfo{eid}{021502}
  (\bibinfo{year}{2008}), \eprint{0709.1915}.

\bibitem[{\citenamefont{{Vines} et~al.}(2011)\citenamefont{{Vines}, {Flanagan},
  and {Hinderer}}}]{Vines}
\bibinfo{author}{\bibfnamefont{J.}~\bibnamefont{{Vines}}},
  \bibinfo{author}{\bibfnamefont{{\'E}.~{\'E}.} \bibnamefont{{Flanagan}}},
  \bibnamefont{and}
  \bibinfo{author}{\bibfnamefont{T.}~\bibnamefont{{Hinderer}}},
  \bibinfo{journal}{Phys. Rev.} \textbf{\bibinfo{volume}{D83}},
  \bibinfo{pages}{084051} (\bibinfo{year}{2011}), \eprint{1101.1673}.

\bibitem[{\citenamefont{Cardoso et~al.}(2017)\citenamefont{Cardoso, Franzin,
  Maselli, Pani, and Raposo}}]{Cardoso:2017cfl}
\bibinfo{author}{\bibfnamefont{V.}~\bibnamefont{Cardoso}},
  \bibinfo{author}{\bibfnamefont{E.}~\bibnamefont{Franzin}},
  \bibinfo{author}{\bibfnamefont{A.}~\bibnamefont{Maselli}},
  \bibinfo{author}{\bibfnamefont{P.}~\bibnamefont{Pani}}, \bibnamefont{and}
  \bibinfo{author}{\bibfnamefont{G.}~\bibnamefont{Raposo}}
  (\bibinfo{year}{2017}), \eprint{1701.01116}.

\bibitem[{\citenamefont{Sennett et~al.}(2017)\citenamefont{Sennett, Hinderer,
  Steinhoff, Buonanno, and Ossokine}}]{Sennett:2017etc}
\bibinfo{author}{\bibfnamefont{N.}~\bibnamefont{Sennett}},
  \bibinfo{author}{\bibfnamefont{T.}~\bibnamefont{Hinderer}},
  \bibinfo{author}{\bibfnamefont{J.}~\bibnamefont{Steinhoff}},
  \bibinfo{author}{\bibfnamefont{A.}~\bibnamefont{Buonanno}}, \bibnamefont{and}
  \bibinfo{author}{\bibfnamefont{S.}~\bibnamefont{Ossokine}},
  \bibinfo{journal}{Phys. Rev.} \textbf{\bibinfo{volume}{D96}},
  \bibinfo{pages}{024002} (\bibinfo{year}{2017}), \eprint{1704.08651}.

\bibitem[{\citenamefont{Johnson-Mcdaniel
  et~al.}(2018)\citenamefont{Johnson-Mcdaniel, Mukherjee, Kashyap, Ajith,
  Del~Pozzo, and Vitale}}]{Johnson-McDaniel2018}
\bibinfo{author}{\bibfnamefont{N.~K.} \bibnamefont{Johnson-Mcdaniel}},
  \bibinfo{author}{\bibfnamefont{A.}~\bibnamefont{Mukherjee}},
  \bibinfo{author}{\bibfnamefont{R.}~\bibnamefont{Kashyap}},
  \bibinfo{author}{\bibfnamefont{P.}~\bibnamefont{Ajith}},
  \bibinfo{author}{\bibfnamefont{W.}~\bibnamefont{Del~Pozzo}},
  \bibnamefont{and} \bibinfo{author}{\bibfnamefont{S.}~\bibnamefont{Vitale}}
  (\bibinfo{year}{2018}), \eprint{1804.08026}.

\bibitem[{\citenamefont{{Vishveshwara}}(1970)}]{VishuNature}
\bibinfo{author}{\bibfnamefont{C.~V.} \bibnamefont{{Vishveshwara}}},
  \bibinfo{journal}{nature} \textbf{\bibinfo{volume}{227}},
  \bibinfo{pages}{936} (\bibinfo{year}{1970}).

\bibitem[{\citenamefont{Dreyer et~al.}(2004)\citenamefont{Dreyer, Kelly,
  Krishnan, Finn, Garrison, and Lopez-Aleman}}]{BHspect04}
\bibinfo{author}{\bibfnamefont{O.}~\bibnamefont{Dreyer}},
  \bibinfo{author}{\bibfnamefont{B.}~\bibnamefont{Kelly}},
  \bibinfo{author}{\bibfnamefont{B.}~\bibnamefont{Krishnan}},
  \bibinfo{author}{\bibfnamefont{L.~S.} \bibnamefont{Finn}},
  \bibinfo{author}{\bibfnamefont{D.}~\bibnamefont{Garrison}}, \bibnamefont{and}
  \bibinfo{author}{\bibfnamefont{R.}~\bibnamefont{Lopez-Aleman}},
  \bibinfo{journal}{Class. Quantum Grav.} \textbf{\bibinfo{volume}{21}},
  \bibinfo{pages}{787} (\bibinfo{year}{2004}), \eprint{gr-qc/0309007}.

\bibitem[{\citenamefont{Berti et~al.}(2009)\citenamefont{Berti, Cardoso, and
  Starinets}}]{Berti:2009kk}
\bibinfo{author}{\bibfnamefont{E.}~\bibnamefont{Berti}},
  \bibinfo{author}{\bibfnamefont{V.}~\bibnamefont{Cardoso}}, \bibnamefont{and}
  \bibinfo{author}{\bibfnamefont{A.~O.} \bibnamefont{Starinets}},
  \bibinfo{journal}{Class. Quant. Grav.} \textbf{\bibinfo{volume}{26}},
  \bibinfo{pages}{163001} (\bibinfo{year}{2009}), \eprint{0905.2975}.

\bibitem[{\citenamefont{Macedo et~al.}(2016)\citenamefont{Macedo, Cardoso,
  Crispino, and Pani}}]{Macedo:2016wgh}
\bibinfo{author}{\bibfnamefont{C.~F.~B.} \bibnamefont{Macedo}},
  \bibinfo{author}{\bibfnamefont{V.}~\bibnamefont{Cardoso}},
  \bibinfo{author}{\bibfnamefont{L.~C.~B.} \bibnamefont{Crispino}},
  \bibnamefont{and} \bibinfo{author}{\bibfnamefont{P.}~\bibnamefont{Pani}},
  \bibinfo{journal}{Phys. Rev.} \textbf{\bibinfo{volume}{D93}},
  \bibinfo{pages}{064053} (\bibinfo{year}{2016}), \eprint{1603.02095}.

\bibitem[{\citenamefont{Berti and Cardoso}(2006)}]{BertiBS}
\bibinfo{author}{\bibfnamefont{E.}~\bibnamefont{Berti}} \bibnamefont{and}
  \bibinfo{author}{\bibfnamefont{V.}~\bibnamefont{Cardoso}},
  \bibinfo{journal}{Int. J. Mod. Phys.} \textbf{\bibinfo{volume}{D15}},
  \bibinfo{pages}{2209} (\bibinfo{year}{2006}), \eprint{gr-qc/0605101}.

\bibitem[{\citenamefont{Macedo et~al.}(2013)\citenamefont{Macedo, Pani,
  Cardoso, and Crispino}}]{PhysRevD.88.064046}
\bibinfo{author}{\bibfnamefont{C.~F.~B.} \bibnamefont{Macedo}},
  \bibinfo{author}{\bibfnamefont{P.}~\bibnamefont{Pani}},
  \bibinfo{author}{\bibfnamefont{V.}~\bibnamefont{Cardoso}}, \bibnamefont{and}
  \bibinfo{author}{\bibfnamefont{L.~C.~B.} \bibnamefont{Crispino}},
  \bibinfo{journal}{Phys. Rev.} \textbf{\bibinfo{volume}{D88}},
  \bibinfo{pages}{064046} (\bibinfo{year}{2013}), \eprint{1307.4812}.

\bibitem[{\citenamefont{Chirenti and
  Rezzolla}(2007{\natexlab{a}})}]{0264-9381-24-16-013}
\bibinfo{author}{\bibfnamefont{C.~B. M.~H.} \bibnamefont{Chirenti}}
  \bibnamefont{and} \bibinfo{author}{\bibfnamefont{L.}~\bibnamefont{Rezzolla}},
  \bibinfo{journal}{Classical and Quantum Gravity}
  \textbf{\bibinfo{volume}{24}}, \bibinfo{pages}{4191}
  (\bibinfo{year}{2007}{\natexlab{a}}).

\bibitem[{\citenamefont{{Pani} et~al.}(2009)\citenamefont{{Pani}, {Berti},
  {Cardoso}, {Chen}, and {Norte}}}]{QNMsGSsPaniAxialPolar}
\bibinfo{author}{\bibfnamefont{P.}~\bibnamefont{{Pani}}},
  \bibinfo{author}{\bibfnamefont{E.}~\bibnamefont{{Berti}}},
  \bibinfo{author}{\bibfnamefont{V.}~\bibnamefont{{Cardoso}}},
  \bibinfo{author}{\bibfnamefont{Y.}~\bibnamefont{{Chen}}}, \bibnamefont{and}
  \bibinfo{author}{\bibfnamefont{R.}~\bibnamefont{{Norte}}},
  \bibinfo{journal}{\prd} \textbf{\bibinfo{volume}{80}}, \bibinfo{eid}{124047}
  (\bibinfo{year}{2009}), \eprint{0909.0287}.

\bibitem[{\citenamefont{Chirenti and
  Rezzolla}(2007{\natexlab{b}})}]{QNMsGSsChirenti2017}
\bibinfo{author}{\bibfnamefont{C.~B. M.~H.} \bibnamefont{Chirenti}}
  \bibnamefont{and} \bibinfo{author}{\bibfnamefont{L.}~\bibnamefont{Rezzolla}},
  \bibinfo{journal}{Class. Quant. Grav.} \textbf{\bibinfo{volume}{24}},
  \bibinfo{pages}{4191} (\bibinfo{year}{2007}{\natexlab{b}}),
  \eprint{0706.1513}.

\bibitem[{\citenamefont{Hartle}(1973)}]{Hartle:1973zz}
\bibinfo{author}{\bibfnamefont{J.~B.} \bibnamefont{Hartle}},
  \bibinfo{journal}{Phys. Rev.} \textbf{\bibinfo{volume}{D8}},
  \bibinfo{pages}{1010} (\bibinfo{year}{1973}).

\bibitem[{\citenamefont{Chatziioannou et~al.}(2013)\citenamefont{Chatziioannou,
  Poisson, and Yunes}}]{Katerina2012}
\bibinfo{author}{\bibfnamefont{K.}~\bibnamefont{Chatziioannou}},
  \bibinfo{author}{\bibfnamefont{E.}~\bibnamefont{Poisson}}, \bibnamefont{and}
  \bibinfo{author}{\bibfnamefont{N.}~\bibnamefont{Yunes}},
  \bibinfo{journal}{Phys. Rev.} \textbf{\bibinfo{volume}{D87}},
  \bibinfo{pages}{044022} (\bibinfo{year}{2013}), \eprint{1211.1686}.

\bibitem[{\citenamefont{Chatziioannou et~al.}(2016)\citenamefont{Chatziioannou,
  Poisson, and Yunes}}]{Katerina2016}
\bibinfo{author}{\bibfnamefont{K.}~\bibnamefont{Chatziioannou}},
  \bibinfo{author}{\bibfnamefont{E.}~\bibnamefont{Poisson}}, \bibnamefont{and}
  \bibinfo{author}{\bibfnamefont{N.}~\bibnamefont{Yunes}},
  \bibinfo{journal}{Phys. Rev.} \textbf{\bibinfo{volume}{D94}},
  \bibinfo{pages}{084043} (\bibinfo{year}{2016}), \eprint{1608.02899}.

\bibitem[{\citenamefont{Maselli et~al.}(2017)\citenamefont{Maselli, Pani,
  Cardoso, Abdelsalhin, Gualtieri, and Ferrari}}]{Maselli:2017cmm}
\bibinfo{author}{\bibfnamefont{A.}~\bibnamefont{Maselli}},
  \bibinfo{author}{\bibfnamefont{P.}~\bibnamefont{Pani}},
  \bibinfo{author}{\bibfnamefont{V.}~\bibnamefont{Cardoso}},
  \bibinfo{author}{\bibfnamefont{T.}~\bibnamefont{Abdelsalhin}},
  \bibinfo{author}{\bibfnamefont{L.}~\bibnamefont{Gualtieri}},
  \bibnamefont{and} \bibinfo{author}{\bibfnamefont{V.}~\bibnamefont{Ferrari}}
  (\bibinfo{year}{2017}), \eprint{1703.10612}.

\bibitem[{\citenamefont{Ryan}(1997{\natexlab{a}})}]{PhysRevD.56.1845}
\bibinfo{author}{\bibfnamefont{F.~D.} \bibnamefont{Ryan}},
  \bibinfo{journal}{Phys. Rev.} \textbf{\bibinfo{volume}{D56}},
  \bibinfo{pages}{1845} (\bibinfo{year}{1997}{\natexlab{a}}).

\bibitem[{\citenamefont{Rodriguez et~al.}(2012)\citenamefont{Rodriguez, Mandel,
  and Gair}}]{Rodriguez:2011aa}
\bibinfo{author}{\bibfnamefont{C.~L.} \bibnamefont{Rodriguez}},
  \bibinfo{author}{\bibfnamefont{I.}~\bibnamefont{Mandel}}, \bibnamefont{and}
  \bibinfo{author}{\bibfnamefont{J.~R.} \bibnamefont{Gair}},
  \bibinfo{journal}{Phys. Rev.} \textbf{\bibinfo{volume}{D85}},
  \bibinfo{pages}{062002} (\bibinfo{year}{2012}), \eprint{1112.1404}.

\bibitem[{\citenamefont{Brown et~al.}(2007)\citenamefont{Brown, Fang, Gair, Li,
  Lovelace, Mandel, and Thorne}}]{Brown:2006pj}
\bibinfo{author}{\bibfnamefont{D.~A.} \bibnamefont{Brown}},
  \bibinfo{author}{\bibfnamefont{H.}~\bibnamefont{Fang}},
  \bibinfo{author}{\bibfnamefont{J.~R.} \bibnamefont{Gair}},
  \bibinfo{author}{\bibfnamefont{C.}~\bibnamefont{Li}},
  \bibinfo{author}{\bibfnamefont{G.}~\bibnamefont{Lovelace}},
  \bibinfo{author}{\bibfnamefont{I.}~\bibnamefont{Mandel}}, \bibnamefont{and}
  \bibinfo{author}{\bibfnamefont{K.~S.} \bibnamefont{Thorne}},
  \bibinfo{journal}{Phys. Rev. Lett.} \textbf{\bibinfo{volume}{99}},
  \bibinfo{pages}{201102} (\bibinfo{year}{2007}), \eprint{gr-qc/0612060}.

\bibitem[{\citenamefont{{Collins} and {Hughes}}(2004)}]{2004PhRvD..69l4022C}
\bibinfo{author}{\bibfnamefont{N.~A.} \bibnamefont{{Collins}}}
  \bibnamefont{and} \bibinfo{author}{\bibfnamefont{S.~A.}
  \bibnamefont{{Hughes}}}, \bibinfo{journal}{\prd}
  \textbf{\bibinfo{volume}{69}}, \bibinfo{eid}{124022} (\bibinfo{year}{2004}),
  \eprint{gr-qc/0402063}.

\bibitem[{\citenamefont{{Glampedakis} and {Babak}}(2006)}]{2006CQGra..23.4167G}
\bibinfo{author}{\bibfnamefont{K.}~\bibnamefont{{Glampedakis}}}
  \bibnamefont{and} \bibinfo{author}{\bibfnamefont{S.}~\bibnamefont{{Babak}}},
  \bibinfo{journal}{Classical and Quantum Gravity}
  \textbf{\bibinfo{volume}{23}}, \bibinfo{pages}{4167} (\bibinfo{year}{2006}),
  \eprint{gr-qc/0510057}.

\bibitem[{\citenamefont{{Ghosh} et~al.}(2018)\citenamefont{{Ghosh},
  {Johnson-McDaniel}, {Ghosh}, {Kant Mishra}, {Ajith}, {Del Pozzo}, {Berry},
  {Nielsen}, and {London}}}]{2018CQGra..35a4002G}
\bibinfo{author}{\bibfnamefont{A.}~\bibnamefont{{Ghosh}}},
  \bibinfo{author}{\bibfnamefont{N.~K.} \bibnamefont{{Johnson-McDaniel}}},
  \bibinfo{author}{\bibfnamefont{A.}~\bibnamefont{{Ghosh}}},
  \bibinfo{author}{\bibfnamefont{C.}~\bibnamefont{{Kant Mishra}}},
  \bibinfo{author}{\bibfnamefont{P.}~\bibnamefont{{Ajith}}},
  \bibinfo{author}{\bibfnamefont{W.}~\bibnamefont{{Del Pozzo}}},
  \bibinfo{author}{\bibfnamefont{C.~P.~L.} \bibnamefont{{Berry}}},
  \bibinfo{author}{\bibfnamefont{A.~B.} \bibnamefont{{Nielsen}}},
  \bibnamefont{and} \bibinfo{author}{\bibfnamefont{L.}~\bibnamefont{{London}}},
  \bibinfo{journal}{Classical and Quantum Gravity}
  \textbf{\bibinfo{volume}{35}}, \bibinfo{eid}{014002} (\bibinfo{year}{2018}),
  \eprint{1704.06784}.

\bibitem[{\citenamefont{Healy et~al.}(2014)\citenamefont{Healy, Lousto, and
  Zlochower}}]{PhysRevD.90.104004}
\bibinfo{author}{\bibfnamefont{J.}~\bibnamefont{Healy}},
  \bibinfo{author}{\bibfnamefont{C.~O.} \bibnamefont{Lousto}},
  \bibnamefont{and}
  \bibinfo{author}{\bibfnamefont{Y.}~\bibnamefont{Zlochower}},
  \bibinfo{journal}{Phys. Rev.} \textbf{\bibinfo{volume}{D90}},
  \bibinfo{pages}{104004} (\bibinfo{year}{2014}), \eprint{1406.7295}.

\bibitem[{\citenamefont{Krishnendu et~al.}(2017)\citenamefont{Krishnendu, Arun,
  and Mishra}}]{Krishnendu:2017shb}
\bibinfo{author}{\bibfnamefont{N.~V.} \bibnamefont{Krishnendu}},
  \bibinfo{author}{\bibfnamefont{K.~G.} \bibnamefont{Arun}}, \bibnamefont{and}
  \bibinfo{author}{\bibfnamefont{C.~K.} \bibnamefont{Mishra}},
  \bibinfo{journal}{Phys. Rev. Lett.} \textbf{\bibinfo{volume}{119}},
  \bibinfo{pages}{091101} (\bibinfo{year}{2017}), \eprint{1701.06318}.

\bibitem[{\citenamefont{Laarakkers and Poisson}(1999)}]{Laarakkers:1997hb}
\bibinfo{author}{\bibfnamefont{W.~G.} \bibnamefont{Laarakkers}}
  \bibnamefont{and} \bibinfo{author}{\bibfnamefont{E.}~\bibnamefont{Poisson}},
  \bibinfo{journal}{Astrophys. J.} \textbf{\bibinfo{volume}{512}},
  \bibinfo{pages}{282} (\bibinfo{year}{1999}), \eprint{gr-qc/9709033}.

\bibitem[{\citenamefont{Pappas and Apostolatos}(2012)}]{Pappas:2012qg}
\bibinfo{author}{\bibfnamefont{G.}~\bibnamefont{Pappas}} \bibnamefont{and}
  \bibinfo{author}{\bibfnamefont{T.~A.} \bibnamefont{Apostolatos}}
  (\bibinfo{year}{2012}), \eprint{1211.6299}.

\bibitem[{\citenamefont{{Pappas} and
  {Apostolatos}}(2012)}]{2012PhRvL.108w1104P}
\bibinfo{author}{\bibfnamefont{G.}~\bibnamefont{{Pappas}}} \bibnamefont{and}
  \bibinfo{author}{\bibfnamefont{T.~A.} \bibnamefont{{Apostolatos}}},
  \bibinfo{journal}{Physical Review Letters} \textbf{\bibinfo{volume}{108}},
  \bibinfo{eid}{231104} (\bibinfo{year}{2012}), \eprint{1201.6067}.

\bibitem[{\citenamefont{Ryan}(1997{\natexlab{b}})}]{Ryan97b}
\bibinfo{author}{\bibfnamefont{F.~D.} \bibnamefont{Ryan}},
  \bibinfo{journal}{Phys. Rev. D} \textbf{\bibinfo{volume}{55}},
  \bibinfo{pages}{6081} (\bibinfo{year}{1997}{\natexlab{b}}).

\bibitem[{\citenamefont{{Dwyer} et~al.}(2015)\citenamefont{{Dwyer}, {Sigg},
  {Ballmer}, {Barsotti}, {Mavalvala}, and {Evans}}}]{CEDwyer}
\bibinfo{author}{\bibfnamefont{S.}~\bibnamefont{{Dwyer}}},
  \bibinfo{author}{\bibfnamefont{D.}~\bibnamefont{{Sigg}}},
  \bibinfo{author}{\bibfnamefont{S.~W.} \bibnamefont{{Ballmer}}},
  \bibinfo{author}{\bibfnamefont{L.}~\bibnamefont{{Barsotti}}},
  \bibinfo{author}{\bibfnamefont{N.}~\bibnamefont{{Mavalvala}}},
  \bibnamefont{and} \bibinfo{author}{\bibfnamefont{M.}~\bibnamefont{{Evans}}},
  \bibinfo{journal}{\prd} \textbf{\bibinfo{volume}{91}}, \bibinfo{eid}{082001}
  (\bibinfo{year}{2015}), \eprint{1410.0612}.

\bibitem[{\citenamefont{Blanchet}(2014)}]{Blanchet:2013haa}
\bibinfo{author}{\bibfnamefont{L.}~\bibnamefont{Blanchet}},
  \bibinfo{journal}{Living Rev. Rel.} \textbf{\bibinfo{volume}{17}},
  \bibinfo{pages}{2} (\bibinfo{year}{2014}), \eprint{1310.1528}.

\bibitem[{\citenamefont{Blanchet et~al.}(2004)\citenamefont{Blanchet, Damour,
  Esposito-Far{\`e}se, and Iyer}}]{BDEI04}
\bibinfo{author}{\bibfnamefont{L.}~\bibnamefont{Blanchet}},
  \bibinfo{author}{\bibfnamefont{T.}~\bibnamefont{Damour}},
  \bibinfo{author}{\bibfnamefont{G.}~\bibnamefont{Esposito-Far{\`e}se}},
  \bibnamefont{and} \bibinfo{author}{\bibfnamefont{B.~R.} \bibnamefont{Iyer}},
  \bibinfo{journal}{Phys. Rev. Lett.} \textbf{\bibinfo{volume}{93}},
  \bibinfo{pages}{091101} (\bibinfo{year}{2004}), \eprint{gr-qc/0406012}.

\bibitem[{\citenamefont{Blanchet et~al.}(2002)\citenamefont{Blanchet, Faye,
  Iyer, and Joguet}}]{BFIJ02}
\bibinfo{author}{\bibfnamefont{L.}~\bibnamefont{Blanchet}},
  \bibinfo{author}{\bibfnamefont{G.}~\bibnamefont{Faye}},
  \bibinfo{author}{\bibfnamefont{B.~R.} \bibnamefont{Iyer}}, \bibnamefont{and}
  \bibinfo{author}{\bibfnamefont{B.}~\bibnamefont{Joguet}},
  \bibinfo{journal}{Phys. Rev. D} \textbf{\bibinfo{volume}{65}},
  \bibinfo{pages}{061501(R)} (\bibinfo{year}{2002}),
  \bibinfo{note}{{Erratum-ibid~{\bf 71}, 129902(E) (2005)}},
  \eprint{gr-qc/0105099}.

\bibitem[{\citenamefont{Blanchet et~al.}(1995)\citenamefont{Blanchet, Damour,
  Iyer, Will, and Wiseman}}]{BDIWW95}
\bibinfo{author}{\bibfnamefont{L.}~\bibnamefont{Blanchet}},
  \bibinfo{author}{\bibfnamefont{T.}~\bibnamefont{Damour}},
  \bibinfo{author}{\bibfnamefont{B.~R.} \bibnamefont{Iyer}},
  \bibinfo{author}{\bibfnamefont{C.~M.} \bibnamefont{Will}}, \bibnamefont{and}
  \bibinfo{author}{\bibfnamefont{A.~G.} \bibnamefont{Wiseman}},
  \bibinfo{journal}{Phys. Rev. Lett.} \textbf{\bibinfo{volume}{74}},
  \bibinfo{pages}{3515} (\bibinfo{year}{1995}), \eprint{gr-qc/9501027}.

\bibitem[{\citenamefont{{Marsat} et~al.}(2013)\citenamefont{{Marsat},
  {Boh{\'e}}, {Faye}, and {Blanchet}}}]{Marsat:2012fn}
\bibinfo{author}{\bibfnamefont{S.}~\bibnamefont{{Marsat}}},
  \bibinfo{author}{\bibfnamefont{A.}~\bibnamefont{{Boh{\'e}}}},
  \bibinfo{author}{\bibfnamefont{G.}~\bibnamefont{{Faye}}}, \bibnamefont{and}
  \bibinfo{author}{\bibfnamefont{L.}~\bibnamefont{{Blanchet}}},
  \bibinfo{journal}{Class. Quant. Grav.} \textbf{\bibinfo{volume}{30}},
  \bibinfo{pages}{055007} (\bibinfo{year}{2013}), \eprint{1210.4143}.

\bibitem[{\citenamefont{{Boh{\'e}} et~al.}(2013)\citenamefont{{Boh{\'e}},
  {Marsat}, {Faye}, and {Blanchet}}}]{Bohe:2012mr}
\bibinfo{author}{\bibfnamefont{A.}~\bibnamefont{{Boh{\'e}}}},
  \bibinfo{author}{\bibfnamefont{S.}~\bibnamefont{{Marsat}}},
  \bibinfo{author}{\bibfnamefont{G.}~\bibnamefont{{Faye}}}, \bibnamefont{and}
  \bibinfo{author}{\bibfnamefont{L.}~\bibnamefont{{Blanchet}}},
  \bibinfo{journal}{Class. Quant. Grav.} \textbf{\bibinfo{volume}{30}},
  \bibinfo{pages}{075017} (\bibinfo{year}{2013}), \eprint{1212.5520}.

\bibitem[{\citenamefont{Boh{\'e} et~al.}(2013)\citenamefont{Boh{\'e}, Marsat,
  and Blanchet}}]{Bohe:2013cla}
\bibinfo{author}{\bibfnamefont{A.}~\bibnamefont{Boh{\'e}}},
  \bibinfo{author}{\bibfnamefont{S.}~\bibnamefont{Marsat}}, \bibnamefont{and}
  \bibinfo{author}{\bibfnamefont{L.}~\bibnamefont{Blanchet}},
  \bibinfo{journal}{Class. Quant. Grav.} \textbf{\bibinfo{volume}{30}},
  \bibinfo{pages}{135009} (\bibinfo{year}{2013}), \eprint{1303.7412}.

\bibitem[{\citenamefont{Marsat et~al.}(2014)\citenamefont{Marsat, Bohé,
  Blanchet, and Buonanno}}]{Marsat:2013caa}
\bibinfo{author}{\bibfnamefont{S.}~\bibnamefont{Marsat}},
  \bibinfo{author}{\bibfnamefont{A.}~\bibnamefont{Bohé}},
  \bibinfo{author}{\bibfnamefont{L.}~\bibnamefont{Blanchet}}, \bibnamefont{and}
  \bibinfo{author}{\bibfnamefont{A.}~\bibnamefont{Buonanno}},
  \bibinfo{journal}{Class. Quant. Grav.} \textbf{\bibinfo{volume}{31}},
  \bibinfo{pages}{025023} (\bibinfo{year}{2014}), \eprint{1307.6793}.

\bibitem[{\citenamefont{Bohé et~al.}(2015)\citenamefont{Bohé, Faye, Marsat,
  and Porter}}]{Bohe:2015ana}
\bibinfo{author}{\bibfnamefont{A.}~\bibnamefont{Bohé}},
  \bibinfo{author}{\bibfnamefont{G.}~\bibnamefont{Faye}},
  \bibinfo{author}{\bibfnamefont{S.}~\bibnamefont{Marsat}}, \bibnamefont{and}
  \bibinfo{author}{\bibfnamefont{E.~K.} \bibnamefont{Porter}},
  \bibinfo{journal}{Class. Quant. Grav.} \textbf{\bibinfo{volume}{32}},
  \bibinfo{pages}{195010} (\bibinfo{year}{2015}), \eprint{1501.01529}.

\bibitem[{\citenamefont{Marsat}(2015)}]{Marsat:2014xea}
\bibinfo{author}{\bibfnamefont{S.}~\bibnamefont{Marsat}},
  \bibinfo{journal}{Class. Quant. Grav.} \textbf{\bibinfo{volume}{32}},
  \bibinfo{pages}{085008} (\bibinfo{year}{2015}), \eprint{1411.4118}.

\bibitem[{\citenamefont{Arun et~al.}(2009)\citenamefont{Arun, Buonanno, Faye,
  and Ochsner}}]{Arun:2008kb}
\bibinfo{author}{\bibfnamefont{K.~G.} \bibnamefont{Arun}},
  \bibinfo{author}{\bibfnamefont{A.}~\bibnamefont{Buonanno}},
  \bibinfo{author}{\bibfnamefont{G.}~\bibnamefont{Faye}}, \bibnamefont{and}
  \bibinfo{author}{\bibfnamefont{E.}~\bibnamefont{Ochsner}},
  \bibinfo{journal}{Phys. Rev.} \textbf{\bibinfo{volume}{D79}},
  \bibinfo{pages}{104023} (\bibinfo{year}{2009}), \bibinfo{note}{[Erratum:
  Phys. Rev.D84,049901(2011)]}, \eprint{0810.5336}.

\bibitem[{\citenamefont{Kidder}(1995)}]{Kidder:1995zr}
\bibinfo{author}{\bibfnamefont{L.~E.} \bibnamefont{Kidder}},
  \bibinfo{journal}{Phys. Rev.} \textbf{\bibinfo{volume}{D52}},
  \bibinfo{pages}{821} (\bibinfo{year}{1995}), \eprint{gr-qc/9506022}.

\bibitem[{\citenamefont{Will and Wiseman}(1996)}]{Will:1996zj}
\bibinfo{author}{\bibfnamefont{C.~M.} \bibnamefont{Will}} \bibnamefont{and}
  \bibinfo{author}{\bibfnamefont{A.~G.} \bibnamefont{Wiseman}},
  \bibinfo{journal}{Phys. Rev.} \textbf{\bibinfo{volume}{D54}},
  \bibinfo{pages}{4813} (\bibinfo{year}{1996}), \eprint{gr-qc/9608012}.

\bibitem[{\citenamefont{Buonanno et~al.}(2013)\citenamefont{Buonanno, Faye, and
  Hinderer}}]{Buonanno:2012rv}
\bibinfo{author}{\bibfnamefont{A.}~\bibnamefont{Buonanno}},
  \bibinfo{author}{\bibfnamefont{G.}~\bibnamefont{Faye}}, \bibnamefont{and}
  \bibinfo{author}{\bibfnamefont{T.}~\bibnamefont{Hinderer}},
  \bibinfo{journal}{Phys. Rev.} \textbf{\bibinfo{volume}{D87}},
  \bibinfo{pages}{044009} (\bibinfo{year}{2013}), \eprint{1209.6349}.

\bibitem[{\citenamefont{Mishra et~al.}(2016)\citenamefont{Mishra, Kela, Arun,
  and Faye}}]{Mishra:2016whh}
\bibinfo{author}{\bibfnamefont{C.~K.} \bibnamefont{Mishra}},
  \bibinfo{author}{\bibfnamefont{A.}~\bibnamefont{Kela}},
  \bibinfo{author}{\bibfnamefont{K.~G.} \bibnamefont{Arun}}, \bibnamefont{and}
  \bibinfo{author}{\bibfnamefont{G.}~\bibnamefont{Faye}}
  (\bibinfo{year}{2016}), \eprint{1601.05588}.

\bibitem[{\citenamefont{Poisson}(1998)}]{Poisson:1997ha}
\bibinfo{author}{\bibfnamefont{E.}~\bibnamefont{Poisson}},
  \bibinfo{journal}{Phys. Rev.} \textbf{\bibinfo{volume}{D57}},
  \bibinfo{pages}{5287} (\bibinfo{year}{1998}), \eprint{gr-qc/9709032}.

\bibitem[{\citenamefont{Mazur and Mottola}(2004)}]{Gravastars}
\bibinfo{author}{\bibfnamefont{P.~O.} \bibnamefont{Mazur}} \bibnamefont{and}
  \bibinfo{author}{\bibfnamefont{E.}~\bibnamefont{Mottola}},
  \bibinfo{journal}{Proc. Nat. Acad. Sci.} \textbf{\bibinfo{volume}{101}},
  \bibinfo{pages}{9545} (\bibinfo{year}{2004}), \eprint{gr-qc/0407075}.

\bibitem[{\citenamefont{Uchikata et~al.}(2016)\citenamefont{Uchikata, Yoshida,
  and Pani}}]{Uchikata:2016qku}
\bibinfo{author}{\bibfnamefont{N.}~\bibnamefont{Uchikata}},
  \bibinfo{author}{\bibfnamefont{S.}~\bibnamefont{Yoshida}}, \bibnamefont{and}
  \bibinfo{author}{\bibfnamefont{P.}~\bibnamefont{Pani}},
  \bibinfo{journal}{Phys. Rev.} \textbf{\bibinfo{volume}{D94}},
  \bibinfo{pages}{064015} (\bibinfo{year}{2016}), \eprint{1607.03593}.

\bibitem[{\citenamefont{Uchikata and Yoshida}(2016)}]{Uchikata:2015yma}
\bibinfo{author}{\bibfnamefont{N.}~\bibnamefont{Uchikata}} \bibnamefont{and}
  \bibinfo{author}{\bibfnamefont{S.}~\bibnamefont{Yoshida}},
  \bibinfo{journal}{Class. Quant. Grav.} \textbf{\bibinfo{volume}{33}},
  \bibinfo{pages}{025005} (\bibinfo{year}{2016}), \eprint{1506.06485}.

\bibitem[{\citenamefont{Cutler and Flanagan}(1994)}]{CutlerFlanagan1994}
\bibinfo{author}{\bibfnamefont{C.}~\bibnamefont{Cutler}} \bibnamefont{and}
  \bibinfo{author}{\bibfnamefont{E.~E.} \bibnamefont{Flanagan}},
  \bibinfo{journal}{Phys. Rev.} \textbf{\bibinfo{volume}{D49}},
  \bibinfo{pages}{2658} (\bibinfo{year}{1994}), \eprint{gr-qc/9402014}.

\bibitem[{\citenamefont{Rao}(1945)}]{Rao45}
\bibinfo{author}{\bibfnamefont{C.}~\bibnamefont{Rao}},
  \bibinfo{journal}{Bullet. Calcutta Math. Soc} \textbf{\bibinfo{volume}{37}},
  \bibinfo{pages}{81} (\bibinfo{year}{1945}).

\bibitem[{\citenamefont{Cramer}(1946)}]{Cramer46}
\bibinfo{author}{\bibfnamefont{H.}~\bibnamefont{Cramer}},
  \emph{\bibinfo{title}{Mathematical methods in statistics}}
  (\bibinfo{publisher}{Pergamon Press}, \bibinfo{address}{Princeton University
  Press, NJ, U.S.A.}, \bibinfo{year}{1946}).

\bibitem[{\citenamefont{Husa et~al.}(2016)\citenamefont{Husa, Khan, Hannam,
  Pürrer, Ohme, Jiménez~Forteza, and Bohé}}]{Husa:2015iqa}
\bibinfo{author}{\bibfnamefont{S.}~\bibnamefont{Husa}},
  \bibinfo{author}{\bibfnamefont{S.}~\bibnamefont{Khan}},
  \bibinfo{author}{\bibfnamefont{M.}~\bibnamefont{Hannam}},
  \bibinfo{author}{\bibfnamefont{M.}~\bibnamefont{Pürrer}},
  \bibinfo{author}{\bibfnamefont{F.}~\bibnamefont{Ohme}},
  \bibinfo{author}{\bibfnamefont{X.}~\bibnamefont{Jiménez~Forteza}},
  \bibnamefont{and} \bibinfo{author}{\bibfnamefont{A.}~\bibnamefont{Bohé}},
  \bibinfo{journal}{Phys. Rev.} \textbf{\bibinfo{volume}{D93}},
  \bibinfo{pages}{044006} (\bibinfo{year}{2016}), \eprint{1508.07250}.

\bibitem[{\citenamefont{Favata et~al.}(In preparation)\citenamefont{Favata,
  Arun, Kim, Kim, and W.Lee}}]{EccPEFavata}
\bibinfo{author}{\bibfnamefont{M.}~\bibnamefont{Favata}},
  \bibinfo{author}{\bibfnamefont{K.~G.} \bibnamefont{Arun}},
  \bibinfo{author}{\bibfnamefont{C.}~\bibnamefont{Kim}},
  \bibinfo{author}{\bibfnamefont{J.}~\bibnamefont{Kim}}, \bibnamefont{and}
  \bibinfo{author}{\bibfnamefont{H.}~\bibnamefont{W.Lee}} (\bibinfo{year}{In
  preparation}).

\bibitem[{\citenamefont{Abbott et~al.}(2016{\natexlab{e}})}]{Evans:2016mbw}
\bibinfo{author}{\bibfnamefont{B.~P.} \bibnamefont{Abbott}}
  \bibnamefont{et~al.} (\bibinfo{collaboration}{LIGO Scientific})
  (\bibinfo{year}{2016}{\natexlab{e}}), \eprint{1607.08697}.

\bibitem[{\citenamefont{Hild et~al.}(2011{\natexlab{b}})\citenamefont{Hild,
  Abernathy, Acernese, Amaro-Seoane, Andersson, Arun, Barone, Barr, Barsuglia,
  Beker et~al.}}]{0264-9381-28-9-094013}
\bibinfo{author}{\bibfnamefont{S.}~\bibnamefont{Hild}},
  \bibinfo{author}{\bibfnamefont{M.}~\bibnamefont{Abernathy}},
  \bibinfo{author}{\bibfnamefont{F.}~\bibnamefont{Acernese}},
  \bibinfo{author}{\bibfnamefont{P.}~\bibnamefont{Amaro-Seoane}},
  \bibinfo{author}{\bibfnamefont{N.}~\bibnamefont{Andersson}},
  \bibinfo{author}{\bibfnamefont{K.}~\bibnamefont{Arun}},
  \bibinfo{author}{\bibfnamefont{F.}~\bibnamefont{Barone}},
  \bibinfo{author}{\bibfnamefont{B.}~\bibnamefont{Barr}},
  \bibinfo{author}{\bibfnamefont{M.}~\bibnamefont{Barsuglia}},
  \bibinfo{author}{\bibfnamefont{M.}~\bibnamefont{Beker}},
  \bibnamefont{et~al.}, \bibinfo{journal}{Classical and Quantum Gravity}
  \textbf{\bibinfo{volume}{28}}, \bibinfo{pages}{094013}
  (\bibinfo{year}{2011}{\natexlab{b}}).

\end{thebibliography}

\end{document}